\documentclass[12pt, draftclsnofoot, onecolumn]{IEEEtran}
\usepackage{epsfig,latexsym}
\usepackage{float}
\usepackage{indentfirst}
\usepackage{amsmath}
\usepackage{amssymb}
\usepackage{times}
\usepackage{subfigure}
\usepackage{psfrag}
\usepackage{cite}
\usepackage{color}
\usepackage{algorithm}
\usepackage{setspace}
\usepackage{mathrsfs}
\usepackage{stfloats}
\usepackage{graphicx}

\begin{document}
\begin{sloppypar}
\title{ \LARGE{Joint Resource Allocation and Configuration Design for STAR-RIS-Enhanced Wireless-Powered MEC }}
\author{Xintong Qin, Zhengyu Song, Tianwei Hou, Wenjuan Yu, Jun Wang, and Xin Sun

\thanks{X. Qin, Z. Song, T. Hou, J. Wang, and X. Sun are with the School of Electronic and Information Engineering, Beijing Jiaotong University, Beijing 100044, China (email: 20111046@bjtu.edu.cn, songzy@bjtu.edu.cn, twhou@bjtu.edu.cn, wangjun1@bjtu.edu.cn, xsun@bjtu.edu.cn).}
\thanks{W. Yu is with the School of Computing and Communications, InfoLab21, Lancaster University, Lancaster LA1 4WA, U.K. (e-mail:  w.yu8@lancaster.ac.uk).}
}

\markboth{}
{Shell \MakeLowercase{\textit{et al.}}: Bare Demo of IEEEtran.cls for Journals}

\maketitle

\begin{abstract}
In this paper, a novel concept called simultaneously transmitting and reflecting RIS (STAR-RIS) is introduced into the wireless-powered mobile edge computing (MEC) systems to improve the efficiency of energy transfer and task offloading. Compared with traditional reflecting-only RIS, STAR-RIS extends the half-space coverage to full-space coverage by simultaneously transmitting and reflecting incident signals, and also provides new degrees-of-freedom (DoFs) for manipulating signal propagation. We aim to maximize the total computation rate of all users, where the energy transfer time, transmit power and CPU frequencies of users, and the configuration design of STAR-RIS are jointly optimized. Considering the characteristics of STAR-RIS, three operating protocols, namely energy splitting (ES), mode switching (MS), and time splitting (TS) are studied, respectively. For the ES protocol, based on the penalty method, successive convex approximation (SCA), and the linear search method, an iterative algorithm is proposed to solve the formulated non-convex problem. Then, the proposed algorithm for ES protocol is extended to solve the MS and TS problems. Simulation results illustrate that the STAR-RIS outperforms traditional reflecting/transmitting-only RIS. More importantly, the TS protocol can achieve the largest computation rate among the three operating protocols of STAR-RIS.
\end{abstract}


\section{Introduction}
\begin{spacing}{1.56}
The wireless communication and Internet of Things (IoT) have advanced at a remarkable pace in the past several years. It is envisioned that the number of IoT user equipments (UEs) will reach around 50 billion by 2030 \cite{UMalik2021}. These resource-limited UEs will generate a large amount of data traffic and require higher communication and computing capacities to meet their demands. As a promising technique, mobile edge computing (MEC) has been regarded as an alternative solution to lift the computing capability of UEs \cite{YMao2017}.

In the MEC systems, the servers are deployed in close proximity (e.g., access point) with UEs and UEs can offload partial or all computation tasks to MEC servers. In so doing, UEs' task execution latency and energy consumption can be effectively reduced \cite{CYou2017} \cite{CXu2015}. Although the MEC can improve the computing performance of UEs, the energy of UEs is another bottleneck for system performance enhancement. Limited by their size, UEs can only store a finite amount of energy. To prevent the depletion of UEs' battery power, the wireless power transfer (WPT) has been regarded as a thriving technology to provide stable and controllable energy supplies for UEs to prolong their lifetime \cite{YAlsaba2018}. By implementing the energy transmitter and energy harvesting modules on the access point (AP) and UEs respectively, UEs can harvest the energy transmitted by the AP, and then utilize the harvested energy to execute their computation tasks by local computing and task offloading \cite{FWang2020}.

Note that both the MEC and WPT are implemented by expanding the functions of the AP, which means these two techniques can be easily integrated to facilitate the wireless-powered MEC \cite{XWang2022}. Such an integration brings many benefits for the system performance enhancement.
For example, when the WPT is introduced into the MEC networks, the UEs can obtain stable energy supplies from the AP via the WPT to execute their tasks \cite{QPham2020}. Meanwhile, thanks to the powerful computation capacity of the MEC server, the task processing time and energy consumption of UEs can be significantly reduced by offloading tasks to the MEC server. More importantly, with less task processing time, more time can be reserved for the UEs’ energy harvesting to prolong their lifetime \cite{YLiu2020}.





However, when the wireless channels between the AP and UEs are blocked by some static or moving obstacles, the efficiency of the energy harvesting and task offloading in the wireless-powered MEC will be greatly reduced. Recently, an emerging paradigm called reconfigurable intelligent surface (RIS) has drawn great attentions due to its capability of smartly reconfiguring the wireless propagation environment \cite{YLiu2021}. The RIS is a man-made and metamaterial-based planar array consisting of a large number of reflecting elements. By dynamically adjusting the phase shift of each element through the smart controller attached to the RIS, the propagation of the wireless signals incident on the RIS can be controlled in a desirable way, such as enhancing the desired signals or eliminating the interference \cite{YLiu2021360}. Nevertheless, most of the existing works assume that the RIS can only reflect the incident signals, which means the transmitter and receiver have to be located in the same side of the RIS. To overcome this geographical restriction, the novel concept of simultaneously transmitting and reflecting RIS (STAR-RIS) has been proposed \cite{JXu2021, XMu2021STAR,YLiuarxiv}

By supporting both the electric and magnetic currents, each element of STAR-RIS can simultaneously reconfigure the transmitted and reflected signals, and thus achieve full-space coverage. Moreover, since both the transmission and reflection coefficient matrices can be designed, the STAR-RIS provides additional degrees-of-freedom (DoFs) to improve the channel conditions. Inspired by the benefits of STAR-RIS, we propose to introduce the STAR-RIS into the wireless-powered MEC.
However, there are still some urgent challenges to be addressed before the STAR-RIS can be harmoniously integrated into the wireless-powered MEC. For example, although the phase shift problem for reflecting-only RIS has been intensively studied in various networks \cite{ZChu2021,XHu2021,ZLi2021}, the proposed algorithms cannot be directly applied to tackle the configuration problem of STAR-RIS. This is because there are three candidate operating protocols for STAR-RIS, namely energy splitting (ES), mode switching (MS), and time splitting (TS) \cite{XMu2021STAR}. The configuration of STAR-RIS needs to be designed according to different operating protocols. Meanwhile, apart from the phase shift, there are more adjustable parameters for the STAR-RIS, i.e., the amplitude adjustments for the ES/MS protocol, and the time allocation for the TS protocol.
These parameters for the STAR-RIS are closely coupled with the communication and computing resource allocation for MEC as well as the energy transfer time for WPT, which results in a highly non-convex optimization problem.
More importantly, despite the existing studies devoted to the optimization for different operating protocols of STAR-RIS in the downlink communications \cite{XMu2021STAR,CWu2022R}, the proposed algorithms is not applicable to the UEs’ task offloading in the uplink. To be specific, when the ES/MS  protocol is employed at the STAR-RIS for the MEC systems, the energy leakage, namely opposite-side leakage appears, where the UEs' uplink offloading energy is wasted since only the transmitted/reflected signals can be received by the AP \cite{ZZhangarxiv}. If the TS protocol is applied by the STAR-RIS, the UEs located in the reflection/transmission space cannot offload task bits when the STAR-RIS operates in the transmission/reflection mode, which results in the reduction of offloading time. Since these operating protocols of STAR-RIS exhibit different features, it is of great importance to compare their performances in the wireless-powered MEC by designing appropriate optimization algorithms.

Motivated by these observations, we investigate the STAR-RIS-enhanced wireless-powered MEC systems in this paper to explore the potential benefits of STAR-RIS on the uplink task offloading and the downlink energy transfer. The main contributions of this paper are summarized as follows.

1) We propose a novel wireless-powered MEC system enhanced by the STAR-RIS, where the mission period is divided into the energy transfer stage and task offloading stage, and the STAR-RIS is deployed to assist the energy transfer in the downlink as well as the task offloading in the uplink. By considering the characteristics of STAR-RIS, the total computation rate maximization problems are formulated for all three operating protocols.

2) In the downlink energy transfer stage, the ES protocol is employed at the STAR-RIS and the coefficient matrices are optimized via a penalty method. Meanwhile, the energy transfer time for the ES/MS and TS protocols is optimized by a linear search method and the linear programming, respectively. In the uplink task offloading stage, the coefficient matrices optimization problems for three operating protocols of STAR-RIS are investigated separately, where an iterative algorithm is proposed to tackle the ES protocol and then the proposed algorithm is extended to solve the MS and TS protocols. Besides, the resource allocation of UEs, i.e., the transmit power and CPU frequency allocation, is optimized based on the successive convex approximation (SCA) technique.

3) Extensive simulation results unveil that for the three operating protocols of STAR-RIS, the computation rate first increases and then decreases with the growth of energy transfer time, while the energy transfer time that results in the maximum computation rate increases with larger mission period. Besides, the STAR-RIS exhibits a better performance than the conventional reflecting/transmitting-only RIS. More importantly, among the three protocols, the TS protocol can achieve the largest computation rate in the proposed system since the transmit power of UEs during the task offloading is able to be fully utilized by the AP and the interference among UEs can be greatly reduced compared to the ES and MS protocols.

The rest of the paper is organized as follows. Related works are discussed in Section II. In Section III, we introduce the system model and formulate the total computation rate maximization problems for all three operating protocols. Section IV elaborates on the proposed algorithms for solving the formulated problems. Some numerical results are shown in Section V, and conclusions are finally drawn in Section VI.
\end{spacing}
\section{Related work}
\subsection{Wireless-Powered MEC}
Recently, the MEC networks integrated with the WPT technique have been extensively investigated \cite{SBi2018,YYu2021,FWang2018}. For example, in \cite{SBi2018}, following the binary offloading policy,
a joint optimization algorithm based on the alternating direction method of multipliers (ADMM) decomposition technique is proposed to maximize the sum computation rate of all users in the wireless-powered MEC, where the users can execute their tasks via local computing or task offloading by utilizing the harvested energy. Differently, benefit from the partial offloading, in \cite{FWang2018}, the AP's energy consumption for computing and WPT, is minimized by optimizing the energy transmit beamforming at the AP, the CPU frequencies and the amount of offloaded bits at the users, as well as the time allocation. As expected, the proposed joint optimization algorithm can significantly reduce the energy consumption of the wireless-powered MEC systems. Then, in \cite{FZhou2020}, subject to the causal constraints for energy harvesting, F. Zhou \emph{et al.} compare the performances of the partial offloading and the binary offloading in the wireless-powered MEC with the aim of maximizing the computation efficiency under the time division multiple access (TDMA) and the non-orthogonal multiple access (NOMA) protocols. Simulation results verify that the partial offloading outperforms the binary offloading and NOMA outperforms TDMA in terms of computation efficiency. Furthermore, in \cite{FWang2020Real}, by considering both the casual task state information (TSI) and channel state information (CSI) for task offloading and energy transfer, the total energy consumption of the wireless-powered MEC system is minimized via jointly optimizing the energy beamforming and the bit allocation, which provides an alternative approach for integrating the WPT into the MEC systems with dynamic task arrivals.
Besides, different from the centralized design of task scheduling and energy charging in the above-mentioned literature, an online learning algorithm is proposed in \cite{XWang2022Online} to minimize the average task completion delay with a distributed execution manner for the wireless-powered MEC networks, where both the simulation results and theoretical analysis demonstrate the advantages of the proposed algorithm in terms of the task completion delay and the convergence speed.

\subsection{RIS and STAR-RIS}
Thanks to the favorable characteristics, RIS has received significant attentions from both the industry and academia. For example, in the RIS-assisted MEC networks \cite{ZChu2021,XHu2021,ZLi2021}, the RIS is deployed to assist the transmission between the MEC server and users by passively reflecting the signals. The channel condition can be improved through optimizing the phase shift of RIS, which is beneficial to the task offloading of users \cite{TBai2020}. Besides, the RIS can also be deployed in the WPT systems. In \cite{LMohjazi2021}, a theoretical framework is developed to investigate the energy sustainability of RIS-assisted WPT systems, where the closed form expression for battery recharging time is obtained when the RIS consists of a large number of elements. Considering the user fairness, in \cite{HYang2021}, the total received power subject to the users' individual minimum received power constraints is maximized by jointly optimizing the beamformer at transmitter and the phase-shifts at the RIS. In \cite{MSun2021}, S. Mao \emph{et al.} deploy the RIS to assist the wirelee-powered MEC, where the total computation bits maximization problem is tackled by an alternative optimization algorithm under the energy casuality constraints of IoT devices and RIS. From the simulation results, it can be seen that the proposed algorithm can achieve higher total computation bits compared to the scheme without RIS.


Despite the attractive features of RIS, the geographical restrictions of transmitter and receiver impose difficulties on the practical implementation, which triggers the emergence of STAR-RIS \cite{CWu2022, THou2022, HNiu2022,HNiu2022Coupled}. Nowadays, the investigation of STAR-RIS is still in its infancy. In \cite{JXu2021}, the concept of STAR-RIS is given, where a general hardware model and two channel models are proposed. Then, the candidate operating protocols of STAR-RIS are investigated in \cite{XMu2021STAR}, which shows the advantages and disadvantages of three protocols of STAR-RIS. Afterwards, in \cite{CWu2022R} and \cite{CWu2021}, the comparison between orthogonal multiple access (OMA) and non-orthogonal multiple access (NOMA) in the STAR-RIS-aided networks is discussed. Numerical results unveil that the integration of NOMA and STAR-RIS significantly outperforms networks employing conventional RIS and OMA. Besides, the STAR-RIS is also deployed to solve the physical-layer security problems. In \cite{HNiu2021} and \cite{YHan2022}, the secrecy rates are maximized and the simulation results show that the secrecy rate of the proposed scheme with STAR-RIS is larger than the conventional RIS, which further verifies the superiority of the STAR-RIS technique.

\section{System Model and Problem Formulation}
\subsection{System Model}
As shown in Fig. 1, we consider a STAR-RIS-enhanced wireless-powered MEC system, which consists of an AP,  multiple UEs indexed by $i \in {\cal I} \buildrel \Delta \over = \left\{ {1,2,...,I} \right\}$, and a STAR-RIS equipped with $M$ passive reflecting/transmitting elements indexed by $m \in{\cal M} = \left\{ {1,2,...,M} \right\}$. According to the location of STAR-RIS, the UE located in the transmission space is denoted by $t \in {\cal T}$, while the UE located in the reflection space is denoted by $r \in {\cal R}$. The numbers of UEs in the reflection and transmission spaces are ${K_r}$ and ${K_t}$, respectively, with ${K_r} + {K_t} = I$, and ${\cal R} \cup {\cal T} = {\cal I}$. In our proposed system, the UEs are equipped with wireless energy-harvesting circuits, communication circuits and computing processors with limited computing capabilities. In addition, the UEs have computation tasks which involve a large amount of task-input data measured by bits. By utilizing the harvested energy, the UEs can execute their task-input data through local computing and task offloading. The AP is endowed with a high performance MEC server to help UEs compute their task-input data. Besides, an RF energy transmitter is also embedded in the AP to provide energy supplies for UEs with the aid of WPT. The STAR-RIS which can simultaneously transmit and reflect the incident signal is deployed to assist the UEs' task offloading and the AP's energy transfer.

\begin{figure}[t]

\centering
\includegraphics[width =3.5in]{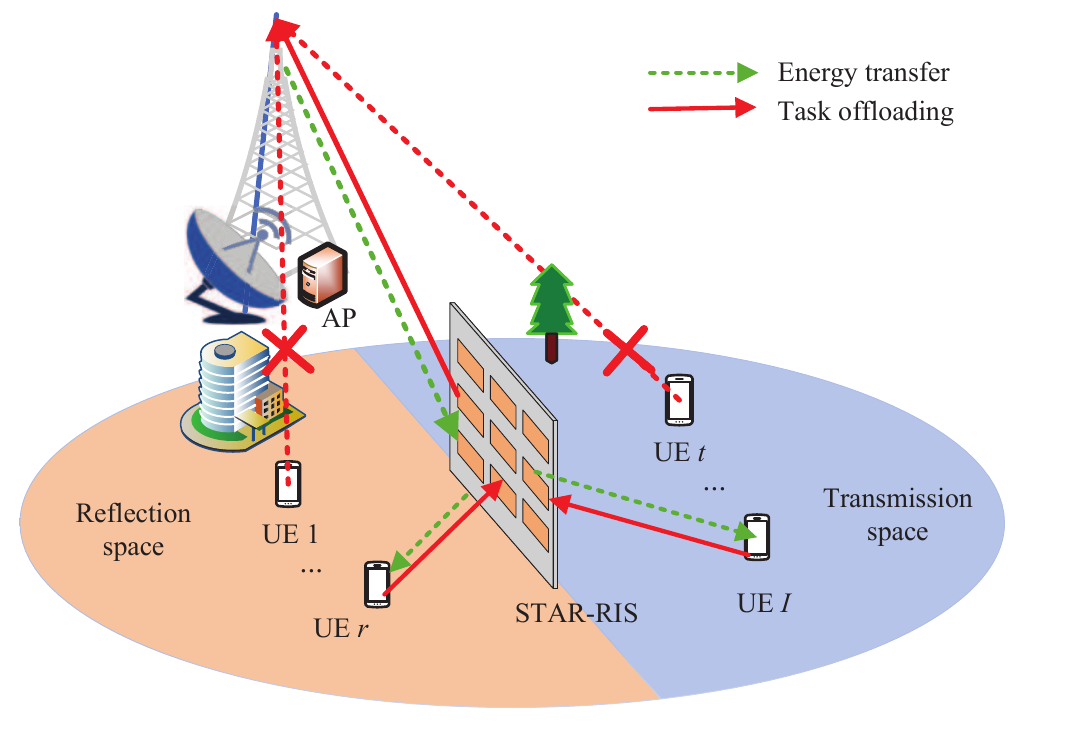} 

\caption{The STAR-RIS-enhanced wireless-powered MEC system.}
\label{3}
\end{figure}

\begin{spacing}{1.56}
\subsubsection{Channel model}
Similar to \cite{XMu2021STAR,CWu2021}, it is assumed that the direct communication links between the AP and the UEs are blocked by obstacles. The downlink channel coefficients from the AP to the STAR-RIS, and from the STAR-RIS to the $i$-th UE are expressed as $g_{{\rm{RIS}}}^{{\rm{AP}}} \in {\mathbb{C}^{M \times 1}}$ and $g_i^{\rm{RIS}} \in {\mathbb{C}^{1 \times M}}$, respectively. The counterpart uplink channel coefficients are given by $h_{\rm{RIS}}^{\rm{AP}} \in {\mathbb{C}^{M \times 1}}$ and $h_i^{\rm{RIS}} \in {\mathbb{C}^{1 \times M}}$. Besides, we suppose that the perfect channel state information of all channels is available at the AP through the advanced channel estimation technologies.

Different from the conventional RIS which can only reflect signals, the incident signal can be divided into the reflected and transmitted signals by each element of the STAR-RIS, thus achieving full-space coverage. There are three protocols for operating the STAR-RIS in the wireless-powered MEC, i.e., ES, MS, and TS \cite{XMu2021STAR,HNiu2022}. To be more specific, for the ES protocol during the task offloading, all elements work in the simultaneous transmission and reflection mode. 
Denoting the amplitude adjustments of the $m$-th element for reflection and transmission as ${\beta _{m,r}^U} $ and $ {\beta _{m,t}^U} $, we have \(\beta _{m,r}^U + \beta _{m,t}^U = 1\) and \(\beta _{m,r}^U,\beta _{m,t}^U \in \left[ {0,1} \right]\).
Note that for the ES protocol, the incident signals from UEs are divided into reflected and transmitted signals by the STAR-RIS, and the AP can only receive the transmitted or reflected signals. Thus, the uplink offloading energy for UEs located in the transmission/reflection space will be leaked to the reflection/transmission side, i.e., the opposite-side leakage appears, which results in the waste of UEs’ offloading energy. Besides, the phase shifts of the $m$-th element for reflection and transmission can be given by $\theta _{m,r}^U$ and $\theta _{m,t}^U$, with $\theta _{m,r}^U,\theta _{m,t}^U \in \left[ {0,2\pi } \right)$. Thus, the reflection- and transmission- coefficient matrices of the STAR-RIS can be given by ${\bf{u}}_{r,{\rm{ES}}}^U = {\rm{diag}}\left( {\sqrt {\beta _{1,r}^U} {e^{j\theta _{1,r}^U}},\sqrt {\beta _{2,r}^U} {e^{j\theta _{2,r}^U}},...,\sqrt {\beta _{M,r}^U} {e^{j\theta _{M,r}^U}}} \right)$ and ${\bf{u}}_{t,{\rm{ES}}}^U = {\rm{diag}}\left( {\sqrt {\beta _{1,t}^U} {e^{j\theta _{1,t}^U}},\sqrt {\beta _{2,t}^U} {e^{j\theta _{2,t}^U}},...,\sqrt {\beta _{M,t}^U} {e^{j\theta _{M,t}^U}}} \right)$.

For the MS protocol, each element of the STAR-RIS can be operated either in reflection mode or transmission mode. Thus, there are binary constraints for the amplitude adjustments, i.e., \(\beta _{m,r}^U + \beta _{m,t}^U = 1\) with \(\beta _{m,r}^U,\beta _{m,t}^U \in \left\{ {0,1} \right\}\). Similar to the ES protocol, the coefficient matrices of MS protocol for STAR-RIS during the task offloading can be expressed as ${\bf{u}}_{r,{\rm{MS}}}^U = {\rm{diag}}\left( {\sqrt {\beta _{1,r}^U} {e^{j\theta _{1,r}^U}},\sqrt {\beta _{2,r}^U} {e^{j\theta _{2,r}^U}},...,\sqrt {\beta _{M,r}^U} {e^{j\theta _{M,r}^U}}} \right)$ and ${\bf{u}}_{t,{\rm{MS}}}^U = {\rm{diag}}\left( {\sqrt {\beta _{1,t}^U} {e^{j\theta _{1,t}^U}},\sqrt {\beta _{2,t}^U} {e^{j\theta _{2,t}^U}},...,\sqrt {\beta _{M,t}^U} {e^{j\theta _{M,t}^U}}} \right)$.

For the TS protocol, all elements of the STAR-RIS work in the same mode (i.e., reflection or transmission mode). Denote the reflection time and transmission time as ${\tau _r}$ and ${\tau _t}$, respectively. By optimizing ${\tau _r}$ and ${\tau _t}$, the STAR-RIS can sequentially switch all elements of STAR-RIS to assist the task offloading of UEs that are located in the reflection and transmission spaces, and hence achieve full-space coverage. Nevertheless, it can be found that the UEs located in the reflection/transmission space cannot offload task bits when the STAR-RIS operates in the transmission/reflection mode, which results in the reduction of offloading time of UEs. The coefficient matrices of TS protocol for reflection and transmission modes are given by ${\bf{u}}_{r,{\rm{TS}}}^U = {\rm{diag}}\left( {{e^{j\theta _{1,r}^U}},{e^{j\theta _{2,r}^U}},...,{e^{j\theta _{M,r}^U}}} \right)$ and ${\bf{u}}_{t,{\rm{TS}}}^U = {\rm{diag}}\left( {{e^{j\theta _{1,t}^U}},{e^{j\theta _{2,t}^U}},...,{e^{j\theta _{M,t}^U}}} \right)$.
\end{spacing}
With the coefficient matrices of STAR-RIS, the combined uplink channel between UE $i$ and the AP can be expressed as \(h_i^U = h_i^{{\rm{RIS}}}{\bf{u}}_{k,X}^Uh_{{\rm{RIS}}}^{{\rm{AP}}}\), where $k \in \left\{ {r,t} \right\}$ represents the operating mode and \(X \in \left\{ {{\rm{ES,MS,TS}}} \right\}\) indicates the employed STAR-RIS operating protocol. If UE $i$ is located in the reflection space, $k = r$. Otherwise, $k = t$.

\subsubsection{Energy transfer and task offloading}
\begin{figure}[t]

\centering
\includegraphics[width =3.5in]{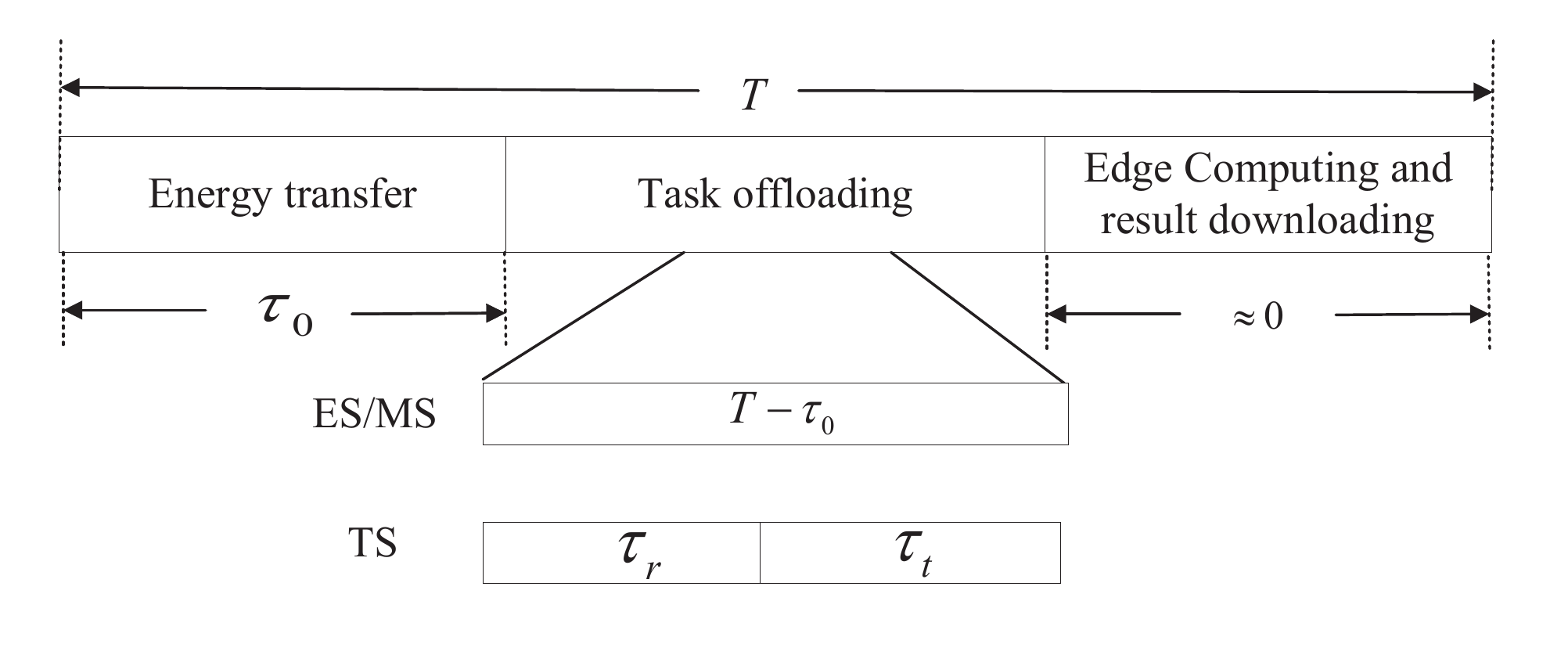}

\caption{The time allocation for UEs in the STAR-RIS-enhanced wireless-powered MEC.}
\label{3}
\end{figure}

As illustrated in Fig. 2, the mission period $T$ is divided into three stages, i.e., the downlink energy transfer stage, the uplink task offloading stage, and the edge computing and result downloading stage. Specifically, in the first stage, the time allocated to the energy transfer is ${\tau _0}$. In order to ensure all UEs in the transmission and reflection spaces can fairly harvest energy, we suppose that the ES protocol is employed at the STAR-RIS for the WPT. Moreover, the process of energy transfer can be regarded as a special multicast transmission. In this case, the ES protocol is appealing since it can make full use of the entire available communication time and allows the UEs to harvest energy all through the first stage \cite{XMu2021STAR}. Thus, the coefficient matrices of STAR-RIS in the downlink energy transfer can be given by ${\bf{u}}_{r,{\rm{ES}}}^D = {\rm{diag}}\left( {\sqrt {\beta _{1,r}^D} {e^{j\theta _{1,r}^D}},\sqrt {\beta _{2,r}^D} {e^{j\theta _{2,r}^D}},...,\sqrt {\beta _{M,r}^D} {e^{j\theta _{M,r}^D}}} \right)$ and ${\bf{u}}_{t,{\rm{ES}}}^D = {\rm{diag}}\left( {\sqrt {\beta _{1,t}^D} {e^{j\theta _{1,t}^D}},\sqrt {\beta _{2,t}^D} {e^{j\theta _{2,t}^D}},...,\sqrt {\beta _{M,t}^D} {e^{j\theta _{M,t}^D}}} \right)$. Similarly, the combined downlink channel between UE $i$ and the AP can be given by \(h_i^D = g_i^{{\rm{RIS}}}{\bf{u}}_{k,{\rm{ES}}}^Dg_{{\rm{RIS}}}^{{\rm{AP}}},k \in \{ r,t\} \).

During the energy transfer stage, the transmit power of the AP is denoted as ${P_0}$. Therefore, the harvested energy of UE $i$ in the first stage can be modeled as \cite{MSun2021}
 \begin{equation}
{P_i} = \eta {\tau _0}{P_0}{\left\| {h_i^D} \right\|^2},
 \end{equation}
where $\eta $ is the energy conversion efficiency for UEs.

In the task offloading stage, the power-domain NOMA is applied and the system bandwidth is denoted as $B$. Since the size of computing result is often trivial compared with that of the original tasks and the computing capacity of MEC server is ultra-high, the time for edge computing and result downloading can be ignored \cite{MSun2021}. Thus, the offloading time of UEs for the ES/MS protocol can be given by $T - {\tau _0}$. While for the TS protocol, the offloading time of UE $i$ can be expressed as
 \begin{equation}
{\tau _{i,k}} = \left\{ {\begin{array}{*{20}{l}}
{{\tau _r},{\rm{if\;UE }}\;i\;{\rm{ is\; located\; in\; the \; reflection \;space;}}}\\
{{\tau _t},{\rm{if\; UE }}\;i\;{\rm{ is\; located\; in\; the \; transmission \;space.}}}
\end{array}} \right.
\end{equation}
Then, denoting the transmit power of UE $i$ and the noise power as ${p_i}$ and ${\sigma ^2}$, respectively, the amount of offloading task bits of UE $i$ for the ES/MS protocol can be expressed as
 \begin{equation}
L_{{\rm{ES/MS}},i}^{{\rm{off}}} = \left( {T - {\tau _0}} \right)B\log \left( {1 + \frac{{{p_i}{{\left\| {h_i^U} \right\|}^2}}}{{\sum\nolimits_{j \ne i}^I {{p_j}{{\left\| {h_j^U} \right\|}^2}}  + {\sigma ^2}}}} \right).
\end{equation}

Correspondingly, if the TS protocol is employed at the STAR-RIS, the amount of offloading task bits of UE $i$ can be given by
 \begin{equation}
L_{{\rm{TS}},i}^{{\rm{off}}} = {\tau _{i,k}}B\log \left( {1 + \frac{{{p_i}{{\left\| {h_i^U} \right\|}^2}}}{{\sum\nolimits_{j \ne i}^I {{p_j}{{\left\| {h_j^U} \right\|}^2}}  + {\sigma ^2}}}} \right).
\end{equation}
\subsubsection{Energy consumption model}
The energy consumption of UEs consists of the energy consumed by task offloading and local computing.
Since the offloading time is different for ES/MS and TS protocols, the offloading energy consumption of UE $i$ in ES/MS can be given by \({p_i}\left( {T - {\tau _0}} \right)\). While for TS, the offloading energy consumption of UE $i$ can be expressed as \({p_i}{\tau _{i,k}}\).

For the local computing, the dynamic voltage and frequency scaling (DVFS) model is adopted to express the UEs' energy consumption \cite{XHu2018,XQin2021}. Denote the CPU frequency of UE $i$ as ${f_i}$. Hence, the energy consumption of UE $i$ for local computing can be given by
 \begin{equation}
E_i^{{\rm{com}}} = \kappa \left( {T - {\tau _0}} \right)f_i^3,
\end{equation}
where $\kappa $ is the effective capacitance coefficient that depends on the processor' s chip architecture.

In addition, the amount of task bits that is computed locally at UE $i$ can be given by
 \begin{equation}
L_i^{{\rm{loc}}} = \frac{{{f_i}\left( {T - {\tau _0}} \right)}}{{{C_i}}},
\end{equation}
where ${C_i}$ is the CPU cycles required for computing 1-bit of task-input data.

\subsection{Problem Formulation}
In this paper, we aim to maximize the total computation rate of all UEs in the STAR-RIS-enhanced wireless-powered MEC system by jointly optimizing the energy transfer time, transmit power and CPU frequencies of UEs, and the configuration design of the STAR-RIS. Based on the characteristics of STAR-RIS, the computation rate maximization problems are formulated for all three operating protocols.
\subsubsection{Problem Formulation for ES/MS Protocol}
The computation rate maximization problems for the ES/MS protocol can be formulated as
  \begin{subequations}
\begin{align}
&\mathop {\max }\limits_{\bf{z}} \sum\limits_{i = 1}^I {\left( {L_i^{{\rm{loc}}} + L_{{\rm{ES/MS,}}i}^{{\rm{off}}}} \right)} \\
{\rm{s}}{\rm{.t}}.&{\rm{ }}\kappa \left( {T \!-\! {\tau _0}} \right)f_i^3\! + \!{p_i}\left( {T \!- \!{\tau _0}} \right) \!\le \!\eta {\tau_0} {P_0}{\left\| {h_i^D} \right\|^2},\forall i \in {\cal I},\\
&{\rm{         }}{f_i} \le {F_{\max }},\forall i \in {\cal I},\\
&{\rm{         }}{p_i} \le {P_{\max }},\forall i \in {\cal I},\\
&{\rm{        }}{\tau _0} \le T,\\
&{\rm{        }}\left\| {\theta _{m,k}^n} \right\| = 1,\forall m \in {\cal M},k \in \{ r,t\} ,n \in \{ D,U\} ,\\
&{\rm{       }}\beta _{m,t}^n + \beta _{m,r}^n = 1,{\rm{ }}\forall m \in {\cal M},n \in \{ D,U\} ,\\
&{\rm{       }}0 \le \beta _{m,t}^n,\beta _{m,r}^n \le 1,\forall m \in {\cal M},n \in \{ D,U\} ,\\
&{\rm{        }}\beta _{m,t}^U,\beta _{m,r}^U \in \{ 0,1\} , \forall m \in {\cal M} \;(\rm{only \;valid\; for \;the \;MS\; protocol}),
\end{align}
 \end{subequations}
where ${\bf{z}} = \left\{ {{\tau _0},{f_i},{p_i},{\bf{u}}_{k,{\rm{ES}}}^D,{\bf{u}}_{k,{\rm{ES/MS}}}^U} \right\}$. ${F_{\max }}$ and ${P_{\max }}$ are the UEs' maximum transmit power and CPU frequency, respectively. Constraint (7b) represents the energy consumption of UE $i$ should be less than the harvested energy from the AP. Constraints (7e) is the feasible set of STAR-RIS's phase shift. (7f) and (7g) are the energy conservation constraints of STAR-RIS. (7i) indicates the binary constraint for each element of STAR-RIS and it is only valid for MS protocol. Note that for the ES protocol, when $\beta _{m,t}^U,\beta _{m,r}^U \in \{ 0,1\} $, it is equivalent to the MS protocol. Thus, from the mathematical perspective, the MS protocol can be regarded as a special case of the ES protocol.
\subsubsection{Problem Formulation for TS Protocol}
 The computation rate maximization problem for the TS protocol can be formulated as
  \begin{subequations}
\begin{align}
&\mathop {\max }\limits_{\bf{z}} \sum\limits_{i = 1}^I {\left( {L_i^{{\rm{loc}}} + L_{{\rm{TS}},i}^{{\rm{off}}}} \right)} \\
{\rm{s}}{\rm{.t}}.&{\rm{    }}\kappa \left( {T - {\tau _0}} \right)f_i^3 + {p_i}{\tau _{i,k}} \le \eta {\tau _0}{P_0}{\left\| {h_i^D} \right\|^2}, \forall i \in {\cal I},\\
&{\rm{         }}{f_i} \le {F_{\max }},\forall i \in {\cal I},\\
&{\rm{         }}{p_i} \le {P_{\max }},\forall i \in {\cal I},\\
&{\rm{        }}\left\| {\theta _{m,k}^n} \right\| = 1,m \in {\cal M},k \in \{ r,t\} ,n \in \{ D,U\} ,\\
&{\rm{        }}\beta _{m,t}^D + \beta _{m,r}^D = 1,m \in {\cal M},\\
&0 \le \beta _{m,t}^D,\beta _{m,r}^D \le 1,m \in {\cal M},\\
&{\rm{        }}{\tau _0} + {\tau _t} + {\tau _r} \le T,
\end{align}
 \end{subequations}
where ${\bf{z}} = \left\{ {{\tau _0},{\tau _r},{\tau _t},{f_i},{p_i},{\bf{u}}_{k,{\rm{ES}}}^D,{\bf{u}}_{k,{\rm{TS}}}^U} \right\}$. Different from ES/MS, there is no uplink energy conservation constraint in TS since all elements of STAR-RIS operate in the same mode, i.e., either transmission or reflection. Constraint (8h) indicates the sum of energy transfer time, the reflection and transmission time of STAR-RIS must be less than the mission period.

\section{Solution to computation rate maximization problems}
In order to tackle the computation rate maximization problems in the STAR-RIS-enhanced wireless-powered MEC system for all three operating protocols, in this section, we first propose an iterative algorithm for the ES protocol based on the SCA technique. Then, the proposed algorithm is extended to solve the computation rate maximization problems for the MS and TS protocols.

\subsection{Solution to the ES Protocol}
Due to the highly-coupled variables and the non-convex objective function, it is difficult to find the globally optimal solution in polynomial time for problem (7) with ES protocol.

\textbf{\emph{Remark 1:}} It can be observed that for any specific energy transfer time ${\tau _0}$, the maximum computation rate can be obtained by optimizing $\left\{ {{p_i},{f_i}} \right\}$ and $\left\{ {{\bf{u}}_{k,{\rm{ES}}}^D,{\bf{u}}_{k,{\rm{ES}}}^U} \right\}$. Inspired by this observation, problem (7) with ES protocol can be decomposed into two subproblems, namely, the resource allocation of UEs and the coefficient matrices optimization for STAR-RIS. Besides, we also note that the optimal energy transfer time ${\tau _0}^*$ cannot be obtained in advance. Fortunately, problem (7) with ES protocol only involves a single continuous variable ${\tau _0}$. Therefore, the linear search method can be executed to obtain the optimal energy transfer time ${\tau _0}^*$.

\subsubsection{Resource allocation of UEs}	
When the energy transfer time ${\tau _0}$ and the coefficient matrices of STAR-RIS $\left\{ {{\bf{u}}_{k,{\rm{ES}}}^D,{\bf{u}}_{k,{\rm{ES}}}^U} \right\}$ are given, problem (7) with ES protocol can be reformulated as
  \begin{subequations}
\begin{align}
&\mathop {\max }\limits_{{p_i},{f_i}} \sum\limits_{i = 1}^I {\left( {L_i^{{\rm{loc}}} + L_{{\rm{ES}},i}^{{\rm{off}}}} \right)} \\
{\rm{s}}{\rm{.t}}.&(7\rm{b})-(7\rm{d}).
\end{align}
 \end{subequations}

To tackle the non-convex objective function in (9), we first define \({R_i} = \log \left( {\sum\nolimits_{j \ne i}^I {{p_j}{{\left\| {h_j^U} \right\|}^2}}  + {\sigma ^2}} \right)\) and thus $L_{{\rm{ES}},i}^{{\rm{off}}}$ can be written as
 \begin{equation}
L_{{\rm{ES}},i}^{{\rm{off}}} = \left( {T - {\tau _0}} \right)B\left( {\log \left( {\sum\limits_{i = 1}^I {{p_i}{{\left\| {h_i^U} \right\|}^2}}  + {\sigma ^2}} \right) - {R_i}} \right).
\end{equation}

By taking the first-order Taylor expansion of ${R_i}$ with respect to ${p_j}$, we have
 \begin{equation}
{R_i} \le \log \left( {\sum\limits_{j \ne i}^I {{p_j}^{(l)}{{\left\| {h_j^U} \right\|}^2}}  + {\sigma ^2}} \right) + \sum\limits_{j \ne i}^I {\frac{{{{\left\| {h_j^U} \right\|}^2}}}{{\ln 2\left( {\sum\nolimits_{j \ne i}^I {{p_j}^{(l)}{{\left\| {h_j^U} \right\|}^2}}  + {\sigma ^2}} \right)}}} \left( {{p_j} - {p_j}^{(l)}} \right) = {\tilde R_i},
\end{equation}
where \({{p_j}^{(l)}}\) is the transmit power of UE $j$ at the $l$-th iteration.

Then, by replacing ${R_i}$ in (10) with ${\tilde R_i}$, during the $l$-th iteration, problem (9) can be approximated as
\begin{subequations}
\begin{align}
&\mathop {\max }\limits_{{p_i},{f_i}} \sum\limits_{i = 1}^I {\left( {T - {\tau _0}} \right)B\left( {\log \left( {\sum\limits_{i = 1}^I {{p_i}{{\left\| {h_i^U} \right\|}^2}}  + {\sigma ^2}} \right) - {{\tilde R}_i}} \right)} + \sum\limits_{i = 1}^I {\frac{{{f_i}\left( {T - {\tau _0}} \right)}}{{{C_i}}}} \\
{\rm{s}}{\rm{.t.}}& (7\rm{b})-(7\rm{d}).
\end{align}
 \end{subequations}

We find that the objective function of (12) is concave with respect to ${f_i}$ and ${p_i}$. Besides, constraint (7b) is convex, and (7c) as well as (7d) are linear. Thus, problem (12) is a convex optimization problem, which can be solved efficiently by standard solvers, such as the CVX \cite{MHua2018}. Then, based on the SCA technique, by iteratively updating $p_i$ and $f_i$ via solving the convex problem (12) until convergence, the solution to problem (9) can be obtained. Thus, the resource allocation algorithm for solving problem (9) with ES protocol can be summarized as Algorithm 1.

\begin{algorithm}[t]
\caption{Resource allocation algorithm for solving problem (9) with ES protocol}
\centering
\begin{tabular}{p{16cm}}
\noindent\hangafter=1\setlength{\hangindent}{1.2em}1. Initialize the vector ${p_i}^{(0)}$ and ${f_i}^{(0)}$, and set the iterative number $l = 0$;

\noindent\hangafter=1\setlength{\hangindent}{1.2em}2. \textbf{while } \(\left| {{{\sum\nolimits_{i = 1}^I {\left( {L_i^{{\rm{loc}}} + L_{{\rm{ES}},i}^{{\rm{off}}}} \right)} }^{(l + 1)}}} \right.\) \(\left. { - {{\sum\nolimits_{i = 1}^I {\left( {L_i^{{\rm{loc}}} + L_{{\rm{ES}},i}^{{\rm{off}}}} \right)} }^{(l)}}} \right| \ge \varepsilon \) \textbf{do }

\noindent\hangafter=1\setlength{\hangindent}{2.4em}3. \hspace{1em} Calculate \({\tilde R_i}^{(l)}\) based on (11);

\noindent\hangafter=1\setlength{\hangindent}{2.4em}4. \hspace{1em} Solve the convex problem (12) to obtain ${p_i}^{(l+1)}$ and ${f_i}^{(l+1)}$;

\noindent\hangafter=1\setlength{\hangindent}{1.2em}5. \hspace{1em} Update the iterative index \(l = l + 1\);

\noindent\hangafter=1\setlength{\hangindent}{1.2em}6. \textbf{end while}
\end{tabular}
\end{algorithm}
\vspace{-0.3in}

\subsubsection{Coefficient Matrices Optimization for STAR-RIS}	
With given ${{\tau _0},{p_i}}$ and ${{f_i}}$, problem (7) with ES protocol can be transformed into
  \begin{subequations}
\begin{align}
&\mathop {\max }\limits_{{\bf{u}}_{k,{\rm{ES}}}^U,{\bf{u}}_{k,{\rm{ES}}}^D} \sum\limits_{i = 1}^I {L_{{\rm{ES}},i}^{{\rm{off}}}} \\
{\rm{s}}{\rm{.t}}{\rm{.}}&(7{\rm{b}}),(7{\rm{e}}) - (7{\rm{g}}).
\end{align}
 \end{subequations}

It can be observed that problem (13) is non-convex and challenging to be solved directly. In order to transform it into a more tractable form, we first define ${h_i} = {\rm{diag}}(h_i^{{\rm{RIS}}})h_{{\rm{RIS}}}^{{\rm{AP}}} \in {\mathbb{C}^{M \times M}}$, ${{\bf{H}}_i} = {h_i}h_i^H$, and ${\bf{V}}_k^U = {\bf{u}}_{k,{\rm{ES}}}^U{\left( {{\bf{u}}_{k,{\rm{ES}}}^U} \right)^H} \in {\mathbb{C}^{M \times M}}$. Thus, the uplink channel gain between UE $i$ and AP can be expressed as ${\left| {h_i^U} \right|^2} = {\left\| {h_i^{{\rm{RIS}}}{\bf{u}}_{k,{\rm{ES}}}^Uh_{{\rm{RIS}}}^{{\rm{AP}}}} \right\|^2} = {\rm{Tr}}({\rm{V}}_k^U{{\bf{H}}_i})$. Similarly, in the downlink, ${\left| {h_i^D} \right|^2} = {\left\| {g_i^{{\rm{RIS}}}{\bf{u}}_{k,{\rm{ES}}}^Dg_{{\rm{RIS}}}^{{\rm{AP}}}} \right\|^2} = {\rm{Tr}}({\bf{V}}_k^D{{\bf{G}}_i})$, where ${\bf{V}}_k^D = {\bf{u}}_{k,{\rm{ES}}}^D{\left( {{\bf{u}}_{k,{\rm{ES}}}^D} \right)^H} \in {\mathbb{C}^{M \times M}}$, ${{\bf{G}}_i} = {g_i}g_i^H$, and ${g_i} = {\rm{diag}}(g_i^{{\rm{RIS}}})g_{{\rm{RIS}}}^{{\rm{AP}}} \in {\mathbb{C}^{M \times M}}$.

Then, problem (13) can be rewritten as
  \begin{subequations}
\begin{align}
&\mathop {\max }\limits_{{\bf{u}}_{k,{\rm{ES}}}^D,{\bf{u}}_{k,{\rm{ES}}}^U} \sum\limits_{i = 1}^I {\left( {T \!-\! {\tau _0}} \right)B\log (1 \!+\! \frac{{{p_i}{\rm{Tr}}({\bf{V}}_k^U{{\bf{H}}_i})}}{{\sum\nolimits_{j \ne i}^I {{p_j}{\rm{Tr}}({\bf{V}}_k^U{{\bf{H}}_j})}  + {\sigma ^2}}})}  \\
&{\rm{s}}{\rm{.t}}{\rm{.    }}\kappa \left( {T - {\tau _0}} \right)f_i^3 + {p_i}\left( {T - {\tau _0}} \right) \le \eta {\tau _0} {P_0}{\mathop{\rm Tr}\nolimits} ({\bf{V}}_k^D{{\bf{G}}_i}),\\
&{\rm{        rank(}}{\bf{V}}_k^n{\rm{) = 1}},n \in \left\{ {D,U} \right\},k \in \left\{ {r,t} \right\},\\
&(7{\rm{e}}) - (7{\rm{g}}),
\end{align}
 \end{subequations}
which is still non-convex due to the objective function and constraints (14b) and (14c). To tackle the non-convex objective function of (14), the auxiliary variables ${{A_i}}$ and ${{B_i}}$ are introduced, with \({1 \mathord{\left/
 {\vphantom {1 {{A_i}}}} \right.
 \kern-\nulldelimiterspace} {{A_i}}} \le {\rm{Tr}}\left( {{\bf{V}}_k^U{{\bf{H}}_i}} \right){p_i}\), and \({B_i} \ge \sum\nolimits_{j \ne i}^I {{\rm{Tr}}\left( {{\bf{V}}_k^U{{\bf{H}}_j}} \right)} {p_j} + {\sigma ^2}\). Then, the objective function can be replaced by  \(\left( {T - {\tau _0}} \right)B\log \left( {1 + {1 \mathord{\left/
 {\vphantom {1 {{A_i}{B_i}}}} \right.
 \kern-\nulldelimiterspace} {{A_i}{B_i}}}} \right)\). By taking the first-order Taylor expansion of $\log \left( {1 + {1 \mathord{\left/
 {\vphantom {1 {{A_i}{B_i}}}} \right.
 \kern-\nulldelimiterspace} {{A_i}{B_i}}}} \right)$ with respect to ${{A_i}}$ and ${{B_i}}$, we have
 \begin{equation}
\begin{array}{l}
\log \left( {1 + \frac{1}{{{A_i}{B_i}}}} \right) \ge \log \left( {1 + \frac{1}{{{A_i}^{(l)}{B_i}^{(l)}}}} \right) - \frac{{\log (e)({A_i} - A_i^{(l)})}}{{A_i^{(l)}(1 + {A_i}^{(l)}{B_i}^{(l)})}} - \frac{{\log (e)({B_i} - B_i^{(l)})}}{{B_i^{(l)}(1 + {A_i}^{(l)}{B_i}^{(l)})}} = {{\hat R}_i},
\end{array}
\end{equation}
where $A_i^{(l)}$ and $B_i^{(l)}$ are local points of $A_i$ and $B_i$ at the $l$-th iteration.
Thus, the non-convex objective function of (14) can be approximated as \(\sum\nolimits_{i = 1}^I {\left( {T - {\tau _0}} \right)B{{\hat R}_i}} \).

\textbf{\emph{Theorem 1:}} When the optimal solution to problem (14) with ES protocol is obtained, constraint (14b) must hold with equality, i.e.,
 \begin{equation}
\kappa \left( {T - {\tau _0}} \right)f_i^3 + {p_i}\left( {T - {\tau _0}} \right) = \eta {\tau _0}{P_0}{\mathop{\rm Tr}\nolimits} ({\bf{V}}_k^D{{\bf{G}}_i}).
\end{equation}

\emph{Proof:} The theorem can be proved by contradiction. Assume that the optimal solution to problem (14) with ES protocol is $\left\{ {{\tau _0}^*,{f_i}^*,{p_i}^*,{\bf{u}}{{_{k,{\rm{ES}}}^D}^*},{\bf{u}}{{_{k,{\rm{ES/MS}}}^U}^*}} \right\}$. If \(\kappa \left( {T - {\tau _0}} \right)f_i^3 + {p_i}\left( {T - {\tau _0}} \right) < \eta {\tau _0}{P_0}{\mathop{\rm Tr}\nolimits} ({\bf{V}}_k^D{{\bf{G}}_i})\), then the following actions can be taken: 1) reduce the energy transfer time ${\tau _0}^*$, and/or 2) increase the local computing frequency $f_i^*$, and/or 3) increase the transmit power $p_i^*$ without violating the other constraints, to further increase the computation rate. Therefore, the assumed optimal solution is not optimal. Thus, the conclusion in Theorem 1 is proved. $ \hfill{}\blacksquare $

According to Theorem 1, the penalty method can be applied to ensure the equality of (14b) and optimize the downlink coefficient matrices of STAR-RIS. To this end, constraint (14b) is transformed into a penalty term added to the objective function.
Thus, problem (14) can be transformed into
  \begin{subequations}
\begin{align}
&\mathop {\max }\limits_{{\bf{u}}_{k,{\rm{ES}}}^D,{\bf{u}}_{k,{\rm{ES}}}^U,A_i,B_i} \sum\limits_{i = 1}^I {\left( {T - {\tau _0}} \right)B{{\hat R}_i}} \notag \\
& + \sum\limits_{i = 1}^I {\mu \left( {\kappa \left( {T - {\tau _0}} \right)f_i^3 + {p_i}\left( {T - {\tau _0}} \right) - \eta {\tau _0}{P_0}{\mathop{\rm Tr}\nolimits} ({\bf{V}}_k^D{{\bf{G}}_i})} \right)} \\
&{\rm{s}}{\rm{.t}}{\rm{.       rank(}}{\bf{V}}_k^n{\rm{) = 1,}}n \in \left\{ {D,U} \right\},k \in \left\{ {r,t} \right\},\\
&{\rm{        }}\frac{1}{{{A_i}}} \le {\rm{Tr}}({\bf{V}}_k^U{{\bf{H}}_i}){p_i}, k \in \left\{ {r,t} \right\},\forall i \in {\cal I},\\
&{\rm{        }}{B_i} \ge \sum\limits_{j \ne i}^I {{\rm{Tr}}({\bf{V}}_k^U{{\bf{H}}_j})} {p_j} + {\sigma ^2}, k \in \left\{ {r,t} \right\},\forall i \in {\cal I},\\
&(7{\rm{e}}) - (7{\rm{g}}).
\end{align}
 \end{subequations}
where $\mu $ is the penalty factor. $\mu $ is chosen as a large positive constant, which can force the penalty term to be equal to zero and then obtain the optimal downlink coefficient matrices of STAR-RIS.

\textbf{\emph{Theorem 2:}} The constraint (18b), i.e., ${\rm{rank}}({\bf{V}}_k^n){\rm{ = 1}}$, can be approximated by ${\rm{Tr}}({\bf{V}}_k^n) - \gamma ({\bf{V}}_k^n,{({\bf{V}}_k^n)^l}) \le \varepsilon $,
where $\gamma ({\bf{V}}_k^n,{({\bf{V}}_k^n)^l}) = {\left\| {{\bf{V}}_k^n} \right\|_s} + \left\langle {\left( {{\bf{V}}_k^n - {{({\bf{V}}_k^n)}^l}} \right),{\partial _{{\bf{V}}_k^n}}{{\left\| {{\bf{V}}_k^n} \right\|}_s}} \right\rangle $ and $\varepsilon $ is a positive threshold.

\emph{Proof:} Denote the $m$-th largest singular value of ${{\bf{V}}_k^n}$ as ${\rho _m}\left( {{\bf{V}}_k^n} \right)$. Thus, we have \({\rm{Tr}}({\bf{V}}_k^n) = \sum\nolimits_{m = 1}^M {{\rho _m}\left( {{\bf{V}}_k^n} \right)} \) and \({\left\| {{\bf{V}}_k^n} \right\|_s} = {\rho _1}\left( {{\bf{V}}_k^n} \right)\), where \({\left\| {{\bf{V}}_k^n} \right\|_s}\) represents the spectral norm of \({{\bf{V}}_k^n}\). When the rank-one constraint is satisfied with \({\rho _1}\left( {{\bf{V}}_k^n} \right) > 0\) and \({\rho _m}\left( {{\bf{V}}_k^n} \right) = 0,m \ne 1\), the rank-one constraint can be transformed into
 \begin{equation}
{\rm{Tr}}\left( {{\bf{V}}_k^n} \right) - {\left\| {{\bf{V}}_k^n} \right\|_s} = 0.
\end{equation}
At the $l$-th iteration, a lower-bound of \({\left\| {{\bf{V}}_k^n} \right\|_s}\) cam be given by
 \begin{equation}
\gamma ({\bf{V}}_k^n,{({\bf{V}}_k^n)^l}) = {\left\| {{\bf{V}}_k^n} \right\|_s} + \left\langle {\left( {{\bf{V}}_k^n - {{({\bf{V}}_k^n)}^l}} \right),{\partial _{{\bf{V}}_k^n}}{{\left\| {{\bf{V}}_k^n} \right\|}_s}} \right\rangle .
\end{equation}
Thus, (19) can be approximated by \({\rm{Tr}}({\bf{V}}_k^n) - \gamma ({\bf{V}}_k^n,{({\bf{V}}_k^n)^l}) \le \varepsilon \). Theorem 2 is proved. $ \hfill{}\blacksquare $

Based on Theorem 2, the rank one constraint (18b) can be tackled by its approximated form. Thus, problem (18) is reformulated as
  \begin{subequations}
\begin{align}
&\mathop {\max }\limits_{{\bf{u}}_{k,{\rm{ES}}}^D,{\bf{u}}_{k,{\rm{ES}}}^U,A_i,B_i} \sum\limits_{i = 1}^I {\left( {T - {\tau _0}} \right)B{{\hat R}_i}} {\rm{ }}\notag\\
 &+ \sum\limits_{i = 1}^I {\mu \left( {\kappa \left( {T - {\tau _0}} \right)f_i^3 + {p_i}\left( {T - {\tau _0}} \right) - \eta {\tau _0}{P_0}{\rm{Tr}}({\bf{V}}_k^D{{\bf{G}}_i})} \right)} \\
&{\rm{s}}.{\rm{t}}.(17{\rm{c}}),(17{\rm{d}}),(7{\rm{e}}) - (7{\rm{g}}),\\
&{\rm{Tr}}({\bf{V}}_k^n) \!-\! \gamma ({\bf{V}}_k^n,{({\bf{V}}_k^n)^l})\! \le\! \varepsilon ,n \in \{ D,U\}, k \in \left\{ {r,t} \right\}.
\end{align}
 \end{subequations}

It can be found that problem (20) is a standard convex semidefinite program (SDP) and can be solved via classic convex optimization toolboxes, such as the SDP solver in CVX \cite{QWu2019}. By iteratively updating ${{A_i},{B_i},}$ and ${{\bf{V}}_k^n}$ via solving the convex problem (20) until convergence, the solution to problem (13) can be achieved and the coefficient matrices of STAR-RIS in both uplink and downlink can be obtained. The proposed SCA-based algorithm for solving problem (13) can be summarized as Algorithm 2.

\begin{algorithm}[t]
\caption{The SCA-based algorithm for solving problem (13) with ES protocol}
\centering
\begin{tabular}{p{16cm}}
\noindent\hangafter=1\setlength{\hangindent}{1.2em}1. Initialize \(\mathop {{\bf{u}}_{k,{\rm{ES}}}^D}\nolimits^{(0)} \), \(\mathop {{\bf{u}}_{k,{\rm{ES}}}^U}\nolimits^{(0)} \), \({A_i}^{(0)}\), and \({B_i}^{(0)}\), and set the iterative number $l = 0$;

\noindent\hangafter=1\setlength{\hangindent}{2.4em}2. \textbf{while} \(\left| {{{\sum\nolimits_{i = 1}^I {L_{{\rm{ES}},i}^{{\rm{off}}}} }^{(l + 1)}} - {{\sum\nolimits_{i = 1}^I {L_{{\rm{ES}},i}^{{\rm{off}}}} }^{(l)}}} \right| \ge \varepsilon \) \textbf{do}

\noindent\hangafter=1\setlength{\hangindent}{2.4em}3. \hspace{1em} Calculate \({\hat R_i}^{(l)}\) and \(\gamma ({\bf{V}}_k^n,{({\bf{V}}_k^n)^l})\) based on (15) and (19), respectively;

\noindent\hangafter=1\setlength{\hangindent}{2.4em}4. \hspace{1em} Solve the convex optimization problem (20) to obtain \(\mathop {{\bf{u}}_{k,{\rm{ES}}}^D}\nolimits^{(l+1)} \), \(\mathop {{\bf{u}}_{k,{\rm{ES}}}^U}\nolimits^{(l+1)} \), \({A_i}^{(l+1)}\), and \({B_i}^{(l+1)}\);

\noindent\hangafter=1\setlength{\hangindent}{1.2em}5. \hspace{1em} Update the iterative index \(l = l + 1\);

\noindent\hangafter=1\setlength{\hangindent}{1.2em}6. \textbf{end while}
\end{tabular}
\end{algorithm}

For the given energy transfer time \({\tau _0}\), the proposed algorithm to solve problem (7) with ES protocol can be outlined in Algorithm 3.

\begin{algorithm}[t]
\caption{Proposed algorithm to solve problem (7) for ES with specific energy transfer time \({\tau _0}\)}
\centering
\begin{tabular}{p{16cm}}
\noindent\hangafter=1\setlength{\hangindent}{1.2em}1. Initialize \({p_i}^{(0)},{f_i}^{(0)}\),\(\mathop {{\bf{u}}_{k,{\rm{ES}}}^D}\nolimits^{(0)} \), \(\mathop {{\bf{u}}_{k,{\rm{ES}}}^U}\nolimits^{(0)} \), and set the iterative number $t = 1$;

\noindent\hangafter=1\setlength{\hangindent}{1.2em}2. \textbf{repeat :}

\noindent\hangafter=1\setlength{\hangindent}{2.4em}3. \hspace{1em} Solve the resource allocation problem (9) to obtain \({p_i}^{(t)}\) and \({f_i}^{(t)}\) by Algorithm 1;

\noindent\hangafter=1\setlength{\hangindent}{2.4em}4. \hspace{1em} Solve the coefficient matrices optimization problem (13) to obtain \(\mathop {{\bf{u}}_{k,{\rm{ES}}}^D}\nolimits^{(t)} \) and \(\mathop {{\bf{u}}_{k,{\rm{ES}}}^U}\nolimits^{(t)} \) by Algorithm 2;

\noindent\hangafter=1\setlength{\hangindent}{2.4em}5. \hspace{1em} Calculate \(L_{{\rm{sum}}}^{{\rm{ES}}}({t}) = \sum\nolimits_{i = 1}^I {\left( {L_i^{{\rm{loc}}} + L_{{\rm{ES}},i}^{{\rm{off}}}} \right)} \);

\noindent\hangafter=1\setlength{\hangindent}{2.4em}6. \hspace{1em} Update the iterative index $t = t + 1$;

\noindent\hangafter=1\setlength{\hangindent}{2.4em}7. \textbf{Until} $t > {N_{\max }}$ or \(\left| {L_{{\rm{sum}}}^{{\rm{ES}}}({t+1}) - L_{{\rm{sum}}}^{{\rm{ES}}}({t})} \right| \le \delta \)

\noindent\hangafter=1\setlength{\hangindent}{1.2em}8. \textbf{Output} the resource allocation and coefficient matrices optimization result \(\left\{ {{p_i}^*,{f_i}^*,{\bf{u}}{{_{k,{\rm{ES}}}^D}^*},{\bf{u}}{{_{k,{\rm{ES}}}^U}^*}} \right\}\).
\end{tabular}
\end{algorithm}

\textbf{\emph{Remark 2:}} Since problem (7) with ES protocol only involves a single continuous variable ${\tau _0}$, the linear search method can be used to obtain the optimal energy transfer time ${\tau _0^*}$. To be specific, within the interval $\left( {0,T} \right)$, ${\tau _0}$ is updated with a small step size $\Delta $.
With given ${\tau _0}$, we solve the total computation rate maximization problems with ES protocol by Algorithm 3 and obtain $\left\{ {{p_i}^*,{f_i}^*,{\bf{u}}{{_{k,{\rm{ES}}}^D}^*},{\bf{u}}{{_{k,{\rm{ES}}}^U}^*}} \right\}$. By examining all the discrete values of ${\tau _0}$, the maximum computation rate and the optimal solution for problem (7) with ES protocol can be obtained with desired accuracy.


Based on Remark 2, the proposed joint resource allocation and coefficient matrices optimization algorithm for solving problem (7) with ES protocol is outlined in Algorithm 4.

\begin{algorithm}[t]
\caption{Joint resource allocation and coefficient matrices optimization algorithm for ES protocol}
\centering
\begin{tabular}{p{16cm}}
\noindent\hangafter=1\setlength{\hangindent}{1.2em}1. Initialize the step size $\Delta $ as a small number and ${\tau _0} \!=\! 0$;

\noindent\hangafter=1\setlength{\hangindent}{1.2em}2. \textbf{while} ${\tau _0} \le T$ \textbf{do:}

\noindent\hangafter=1\setlength{\hangindent}{2.4em}3. \hspace{1em} Run Algorithm 3 to obtain \({p_i}^{*},{f_i}^{*}\),\(\mathop {{\bf{u}}_{k,{\rm{ES}}}^D}\nolimits^{*} \), and \(\mathop {{\bf{u}}_{k,{\rm{ES}}}^U}\nolimits^{*} \);

\noindent\hangafter=1\setlength{\hangindent}{2.4em}4. \hspace{1em} Denote the objective function of (7) as \(L_{{\rm{sum}}}^{{\rm{ES}}}({\tau _0})\);

\noindent\hangafter=1\setlength{\hangindent}{1.2em}5. \hspace{1em} Update \({\tau _0} = {\tau _0} + \Delta \);

\noindent\hangafter=1\setlength{\hangindent}{1.2em}6. \textbf{end while};

\noindent\hangafter=1\setlength{\hangindent}{1.2em}7. \textbf{Output} $\tau _0^* = \arg {\max _{{\tau _0}}}L_{{\rm{sum}}}^{{\rm{ES}}}({\tau _0})$ and the corresponding resource allocation and coefficient matrices optimization result \(\left\{ {{p_i}^*,{f_i}^*,{\bf{u}}{{_{k,{\rm{ES}}}^D}^*}{\rm{ }},{\rm{ }}{\bf{u}}{{_{k,{\rm{ES}}}^U}^*}{\rm{ }}} \right\}\) for problem (7) with ES protocol.
\end{tabular}
\end{algorithm}

\subsection{Solution to the MS Protocol}
Compared to the ES protocol, problem (7) with the MS protocol is a mixed-integer non-convex optimization problem due to the binary constraint (7i). To tackle this problem, the binary constraint is equivalently transformed into an equality, i.e.,
 \begin{equation}
\beta _{m,k}^U\left( {\beta _{m,k}^U - 1} \right) = 0.
\end{equation}

By further adding the equality constraint (22) as another penalty term into the objective function, problem (7) with MS protocol can be transformed into
  \begin{subequations}
\begin{align}
&\mathop {\max }\limits_{\bf{z}} \sum\limits_{i = 1}^I {\left( {L_i^{{\rm{loc}}} + L_{{\rm{MS,}}i}^{{\rm{off}}}} \right)}  + \sum\limits_{m = 1}^M {\sum\limits_{k \in \left\{ {r,t} \right\}} {\nu \left( {\beta _{m,k}^U\left( {\beta _{m,k}^U - 1} \right)} \right)} } \\
&{\rm{s}}{\rm{.t}}{\rm{.(7b) - (7g)}},
\end{align}
 \end{subequations}
where $\nu$ is a positive penalty factor. Similar to the method for ES protocol, with given ${\tau _0}$, the problem is decomposed into two subproblems, namely, the resource allocation of UEs and the coefficient matrices optimization for STAR-RIS with the MS protocol. To be more specific, for the subproblem of resource allocation of UEs, it can be solved by a similar method to the ES protocol. While for the subproblem of coefficient matrices optimization for MS protocol, since the penalty term about \(\beta _{m,k}^U\) renders the objective function non-convex, the Taylor expansion is exploited again to obtain the lower convex bound of the penalty term at the $l$-th iteration, i.e.,
 \begin{equation}
\beta _{m,k}^U\left( {\beta _{m,k}^U - 1} \right) \ge \left( {2\beta {{_{m,k}^U}^{(l)}} - 1} \right)\beta _{m,k}^U - {\left( {\beta _{m,k}^U} \right)^2}.
\end{equation}
Thus, the subproblem of coefficient matrices optimization for MS protocol during the $l$-th iteration can be reformulated as
  \begin{subequations}
\begin{align}
&\mathop {\max }\limits_{{\bf{u}}_{k,{\rm{MS}}}^U,{\bf{u}}_{k,{\rm{ES}}}^D,A_i,B_i} \sum\limits_{i = 1}^I {\left( {T - {\tau _0}} \right)B{{\hat R}_i}} + \sum\limits_{i = 1}^I {\mu \left( {\kappa \left( {T - {\tau _0}} \right)f_i^3 + {p_i}\left( {T - {\tau _0}} \right) - \eta {\tau _0}{P_0}{{\left\| {h_i^D} \right\|}^2}} \right)} \notag\\
 &+ \sum\limits_{m = 1}^M {\sum\limits_{k \in \left\{ {r,t} \right\}} {\nu \left( {(2\beta {{_{m,k}^U}^{(l)}} - 1)\beta _{m,k}^U - {{\left( {\beta _{m,k}^U} \right)}^2}} \right)} } \\
&{\rm{s}}{\rm{.t}}{\rm{.  (20b);(20c)}},
\end{align}
 \end{subequations}
which is a standard convex problem.

\begin{algorithm}[t]
\caption{Proposed algorithm to solve problem (7) with MS protocol }
\centering
\begin{tabular}{p{16cm}}
\noindent\hangafter=1\setlength{\hangindent}{1.2em}1. Initialize the variables
\({p_i}^{(0)},{f_i}^{(0)}\),\(\mathop {{\bf{u}}_{k,{\rm{ES}}}^D}\nolimits^{(0)} \), \(\mathop {{\bf{u}}_{k,{\rm{MS}}}^U}\nolimits^{(0)} \), the small step $\Delta $, and set ${\tau _0} = 0$;

\noindent\hangafter=1\setlength{\hangindent}{1.2em}2. \textbf{repeat :} Outer loop

\noindent\hangafter=1\setlength{\hangindent}{2.4em}3. \hspace{1em} Set the energy transfer time ${\tau _0} = {\tau _0} + \Delta $ and the iteration number $t = 0$;

\noindent\hangafter=1\setlength{\hangindent}{2.4em}4. \hspace{1em} \textbf{repeat :} Inner loop

\noindent\hangafter=1\setlength{\hangindent}{3.6em}5. \hspace{2em} Solve the resource allocation problem for MS to obtain \({p_i}^{(t)}\) and \({f_i}^{(t)}\);

\noindent\hangafter=1\setlength{\hangindent}{3.6em}6. \hspace{2em} Solve the coefficient matrices optimization problem for MS to obtain \(\mathop {{\bf{u}}_{k,{\rm{ES}}}^D}\nolimits^{(t)} \) and \(\mathop {{\bf{u}}_{k,{\rm{MS}}}^U}\nolimits^{(t)} \);

\noindent\hangafter=1\setlength{\hangindent}{3.6em}7.\hspace{2em} Calculate $L_{{\rm{sum}}}^{{\rm{MS}}}({{(t)}}) = \sum\nolimits_{i = 1}^I {\left( {L_i^{{\rm{loc}}} + L_{{\rm{MS}},i}^{{\rm{off}}}} \right)} $;

\noindent\hangafter=1\setlength{\hangindent}{3.6em}8.\hspace{2em} Update the iterative index $t = t + 1$;

\noindent\hangafter=1\setlength{\hangindent}{2.4em}9. \hspace{1em} \textbf{Until} $t > {N_{\max }}$ or \(\left| {L_{{\rm{sum}}}^{{\rm{MS}}}({t+1}) - L_{{\rm{sum}}}^{{\rm{MS}}}({t})} \right| \le \delta \);

\noindent\hangafter=1\setlength{\hangindent}{1.2em}10.  \hspace{1em} Update \(L({\tau _0}) = L_{{\rm{sum}}}^{{\rm{MS}}}({{(t + 1)}})\);

\noindent\hangafter=1\setlength{\hangindent}{1.2em}11. \textbf{Until :} \({\tau _0} > T\);

\noindent\hangafter=1\setlength{\hangindent}{1.2em}12. \textbf{Output :} \(\tau _0^* = \arg {\max _{{\tau _0}}}(L({\tau _0}))\) and the corresponding solution to resource allocation and coefficient matrices optimization \(\left\{ {{p_i}^*,{f_i}^*,{\bf{u}}{{_{k,{\rm{ES}}}^D}^*},{\bf{u}}{{_{k,{\rm{MS}}}^U}^*}} \right\}\).
\end{tabular}
\end{algorithm}
\vspace{0.3in}
By iteratively solving the subproblems of UEs' resource allocation and the coefficient matrices optimization for STAR-RIS with MS protocol, the computation rate maximization problem with given energy transfer time ${\tau _0}$ can be handled effectively. Then, by exploiting the linear search method, the optimal energy transfer time ${\tau _0^*}$ and the corresponding solution to problem (7) with MS protocol can be finally obtained. The proposed algorithm for solving problem (7) with MS protocol is outlined in Algorithm 5.
\vspace{-0.2in}
\subsection{Solution to the TS Protocol}
To solve the total computation rate maximization problem for the TS protocol, problem (8) is decomposed into three subproblems, i.e., the time allocation, the resource allocation of UEs, and the coefficient matrices optimization for STAR-RIS.

\textbf{\emph{Remark 3:}} When the TS protocol is employed at the STAR-RIS during the task offloading, all elements work either in the transmission mode or the reflection mode depending on the transmission and reflection time allocation. Specifically, when the elements of STAR-RIS work in the transmission mode, \(\beta _{m,t}^U = 1,\beta _{m,r}^U = 0\). Otherwise, \(\beta _{m,t}^U = 0,\beta _{m,r}^U = 1\). Thus, there is no need to optimize \(\beta _{m,k}^U\) for the TS protocol. Nevertheless, the time allocation for the TS protocol is more complex compared with the ES/MS protocol, since the time allocated to the energy transfer, the reflection and transmission modes need to be jointly considered. To overcome this difficulty, we first define
 \begin{equation}
\begin{array}{*{20}{l}}
{{\bf{A}} = \left[ { - \sum\limits_{i = 1}^I {\frac{{{f_i}}}{{{C_i}}}\;\;\;\sum\limits_{r = 1}^{{K_r}} {B\log \left( {1 + \frac{{{p_r}{{\left\| {h_r^U} \right\|}^2}}}{{\sum\nolimits_{j \ne r}^{{K_r}} {{p_j}{{\left\| {h_j^U} \right\|}^2}}  + {\sigma ^2}}}} \right)\;\;\;\sum\limits_{t = 1}^{{K_t}} {B\log \left( {1 + \frac{{{p_t}{{\left\| {h_t^U} \right\|}^2}}}{{\sum\nolimits_{j \ne t}^{{K_t}} {{p_j}{{\left\| {h_j^U} \right\|}^2}}  + {\sigma ^2}}}} \right)} } } } \right],}
\end{array}
\end{equation}
 \begin{equation}
{\bf{C}} = \left[ {\begin{array}{*{20}{c}}
{ - \kappa f_r^3 - \eta {P_0}{{\left\| {h_r^D} \right\|}^2}}&{{p_r}}&0\\
{ - \kappa f_t^3 - \eta {P_0}{{\left\| {h_t^D} \right\|}^2}}&0&{{p_t}}\\
1&1&1
\end{array}} \right],
\end{equation}
 \begin{equation}
{\bf{D}} = {\left[ {\begin{array}{*{20}{c}}
{ - \kappa Tf_r^3}&{ - \kappa Tf_t^3}&T
\end{array}} \right]^H}.
 \end{equation}

Therefore, for problem (8), when the results for resource allocation of UEs \(\left\{ {{p_i},{f_i}} \right\}\) and the coefficient matrices optimization of STAR-RIS $\left\{{{\bf{u}}_{k,{\rm{ES}}}^D,{\bf{u}}_{k,{\rm{TS}}}^U} \right\}$ are given, the time allocation problem for the TS protocol can be expressed as
  \begin{subequations}
\begin{align}
&\mathop {\max }\limits_{\bf{\Lambda}}  {\bf{A}} {\bf{\Lambda}} \\
{\rm{s}}{\rm{.t}}{\rm{.}}&{\bf{C}}{\bf{\Lambda}}  \le {\bf{D}},\\
&{\bf{\Lambda}}  \ge 0,
\end{align}
 \end{subequations}
where ${\bf{\Lambda }} = {\left[ {{\tau _0}\;\;{\tau _r}\;\;{\tau _t}} \right]^H}$. (28) is a standard linear programming (LP) problem and can be easily solved.

Then, with given time allocation \({\bf{\Lambda }}\) and the coefficient matrices of STAR-RIS $\left\{{{\bf{u}}_{k,{\rm{ES}}}^D,{\bf{u}}_{k,{\rm{TS}}}^U} \right\}$, the subproblem of UEs' resource allocation for the TS protocol can be formulated as
  \begin{subequations}
\begin{align}
\mathop {\max }\limits_{{p_i},{f_i}} \sum\limits_{i = 1}^I {\left( {L_i^{{\rm{loc}}} + L_{{\rm{TS}},i}^{{\rm{off}}}} \right)} \\
{\rm{s}}{\rm{.t}}{\rm{.(8b) - (8d)}}{\rm{.}}
\end{align}
 \end{subequations}

With given time allocation \({\bf{\Lambda }}\) and the resource allocation of UEs \(\left\{ {{p_i},{f_i}} \right\}\), the subproblem of coefficient matrices optimization for STAR-RIS with the TS protocol can be formulated as
  \begin{subequations}
\begin{align}
\mathop {\max }\limits_{{\bf{u}}_{k,{\rm{TS}}}^U,{\bf{u}}_{k,{\rm{ES}}}^D} \sum\limits_{i = 1}^I {L_{{\rm{TS}},i}^{{\rm{off}}}} \\
{\rm{s}}.{\rm{t}}.(8{\rm{b}}),(8{\rm{e}}) - (8{\rm{g}}).
\end{align}
 \end{subequations}

It can be found that the subproblems of resource allocation of UEs and the coefficient matrices optimization for STAR-RIS with TS protocol have similar structures to the corresponding subproblems (9) and (13) with ES protocol. Therefore, subproblems (29) and (30) can be handled by a similar method to the ES protocol. By iteratively solving subproblems (28)-(30), problem (8) for TS protocol can be solved effectively. Thus, the proposed algorithm for the total computation rate maximization problem with TS protocol can be outlined in Algorithm 6.

\begin{algorithm}[t]
\caption{Proposed algorithm for solving problem (8) with TS protocol}
\centering
\begin{tabular}{p{16cm}}
\noindent\hangafter=1\setlength{\hangindent}{1.2em}1. Initialize the vector ${{\bf{\Lambda}} ^{(0)}}$,\({p_i}^{(0)},{f_i}^{(0)}\),\(\mathop {{\bf{u}}_{k,{\rm{ES}}}^D}\nolimits^{(0)} \), \(\mathop {{\bf{u}}_{k,{\rm{TS}}}^U}\nolimits^{(0)} \) and set the iteration number $t=0$;

\noindent\hangafter=1\setlength{\hangindent}{1.2em}2. \textbf{repeat:}

\noindent\hangafter=1\setlength{\hangindent}{2.4em}3. \hspace{1em} Solve the LP problem (28) for TS to obtain the time allocation ${\bf{\Lambda}}^{(t)} $;

\noindent\hangafter=1\setlength{\hangindent}{2.4em}4. \hspace{1em} Solve the resource allocation problem for TS to obtain \({p_i}^{(t)}\) and \({f_i}^{(t)}\);

\noindent\hangafter=1\setlength{\hangindent}{2.4em}5. \hspace{1em} Solve the coefficient matrices optimization problem for TS to obtain \(\mathop {{\bf{u}}_{k,{\rm{ES}}}^D}\nolimits^{(t)} \) and \(\mathop {{\bf{u}}_{k,{\rm{TS}}}^U}\nolimits^{(t)} \);

\noindent\hangafter=1\setlength{\hangindent}{1.2em}6. \hspace{1em} Calculate $L_{{\rm{sum}}}^{{\rm{TS}}}({{(t)}}) = \sum\nolimits_{i = 1}^I {\left( {L_i^{{\rm{loc}}} + L_{{\rm{TS}},i}^{{\rm{off}}}} \right)} $;

\noindent\hangafter=1\setlength{\hangindent}{1.2em}7. Update the iterative index $t = t + 1$;

\noindent\hangafter=1\setlength{\hangindent}{1.2em}8. \textbf{Until:} $t \!>\! {N_{\max }}$ or \(\left| {L_{{\rm{sum}}}^{{\rm{TS}}}({{(t + 1)}}) \!- \!L_{{\rm{sum}}}^{{\rm{TS}}}({{(t)}})} \right| \le \delta \);

\noindent\hangafter=1\setlength{\hangindent}{1.2em}9. \textbf{Output:} the result to the time allocation, resource allocation, and coefficient matrices optimization \(\left\{ {{{\bf{\Lambda }}^*},{p_i}^*,{f_i}^*,{\bf{u}}{{_{k,{\rm{ES}}}^D}^*},{\bf{u}}{{_{k,{\rm{TS}}}^U}^*}} \right\}\).
\end{tabular}
\end{algorithm}

\subsection{Convergence and Complexity Analysis}
The computational complexity of the proposed Algorithm 4 for ES mainly depends on Algorithm 3. Therefore, we first analyze the computational complexity of Algorithm 3. In Algorithm 3, two subproblems are iteratively solved to obtain $\left\{ {{f_i},{p_i}} \right\}$ and \(\left\{ {{\bf{u}}_{k,{\rm{ES}}}^U,{\bf{u}}_{k,{\rm{ES}}}^D} \right\}\) with given energy transfer time ${{\tau _0}}$. For the UE resource allocation subproblem, it can be solved by the interior point method. Thus, the computational complexity of the $\tilde \varepsilon $ optimal solution can be expressed as \({{\cal O}_1} \buildrel \Delta \over = {\cal O}\left( {{L_1}\ln (1/\tilde \varepsilon ){n^3}} \right)\), where $n = 2I$ is the number of decision variables and ${L_1}$ is the number of iterations. For the coefficient matrices optimization subproblem, it can be solved by the SDP method. Denote the number of iterations as ${L_2}$. The computational complexity can be given by ${{\cal O}_2} \buildrel \Delta \over = {\cal O}\left( {{L_2}\ln (1/\tilde \varepsilon ){{(2M)}^{3.5}}} \right)$ \cite{XMu2021STAR}. Thus, the overall computational complexity of Algorithm 3 can be denoted as ${\cal O}\left( {{L_3}\left( {{{\cal O}_1} + {{\cal O}_2}} \right)} \right)$, where ${L_3}$ is the number of iterations. Then, in order to obtain the maximum total computation rate of UEs, Algorithm 4 requires to execute a linear search of ${\tau _0}$ with a small step size $\Delta $ and run Algorithms 3 for $T/\Delta $ times. Thus, the overall complexity of Algorithm 4 also depends on the step size $\Delta $, but it is always polynomial regardless of $\Delta $.

\textbf{\emph{Theorem 3:}} Algorithm 3 increases the total computation rate of UEs at each iteration and finally converges within a limited number of iterations.

\emph{Proof:} Denote the objective function of problem (7) for ES with given ${\tau _0}$ as $L_{{\rm{sum}}}^{{\rm{ES}}}$. We have
\begin{equation}
\begin{array}{l}
L_{{\rm{sum}}}^{{\rm{ES}}}\left( {p_i^{(t)},f_i^{(t)},{\bf{u}}{{_{k,{\rm{ES}}}^D}^{(t)}},{\bf{u}}{{_{k,{\rm{ES}}}^U}^{(t)}}} \right)
\le L_{{\rm{sum}}}^{{\rm{ES}}}\left( {p_i^{(t + 1)},f_i^{(t + 1)},{\bf{u}}{{_{k,{\rm{ES}}}^D}^{(t)}},{\bf{u}}{{_{k,{\rm{ES}}}^U}^{(t)}}} \right)\\
 \;\;\;\;\;\;\;\;\;\;\;\;\;\;\;\;\;\;\;\;\;\;\;\;\;\;\;\;\;\;\;\;\;\;\;\;\;\;\;\;\;\;\;\;\;\;\;\;\;\; \le L_{{\rm{sum}}}^{{\rm{ES}}}\left( {p_i^{(t + 1)},f_i^{(t + 1)},{\bf{u}}{{_{k,{\rm{ES}}}^D}^{(t + 1)}},{\bf{u}}{{_{k,{\rm{ES}}}^U}^{(t + 1)}}} \right).
\end{array}
\end{equation}

The first inequality holds since for fixed coefficient matrices of STAR-RIS \(\left\{ {{\bf{u}}{{_{k,{\rm{ES}}}^D}^{(t)}},{\bf{u}}{{_{k,{\rm{ES}}}^U}^{(t)}}} \right\}\), the optimal $\left\{ {{p_i}^{(t + 1)},{f_i}^{(t + 1)}} \right\}$ is obtained by solving problem (9) via Algorithm 1; the second inequality follows the fact that the optimal \(\left\{ {{\bf{u}}{{_{k,{\rm{ES}}}^D}^{(t + 1)}},{\bf{u}}{{_{k,{\rm{ES}}}^U}^{(t + 1)}}} \right\}\) is obtained by solving problem (13) via Algorithm 2 with given $\left\{ {{p_i}^{(t + 1)},{f_i}^{(t + 1)}} \right\}$. Thus, the objective function of problem (7) for the ES protocol is monotonically non-decreasing after each iteration. In addition, since the total computation rate is upper-bounded, the objective function $L_{{\rm{sum}}}^{{\rm{ES}}}$ must converge after several iterations.  $ \hfill{}\blacksquare $

Algorithm 4 can be regarded as a repetitive execution of Algorithm 3. Therefore, we only need to guarantee the convergence of Algorithm 3, which has been proved in Theorem 3. Since Algorithm 5 has a similar structure to Algorithm 4, its computational complexity and convergence can refer to Algorithm 4.

Different form Algorithm 4, Algorithm 6 consists of three subproblems. For the time allocation subproblem, it can be solved by linear programming method. Thus, the computational complexity can be given by ${\cal O}\left( {{n_1}\ln {n_1}} \right)$, where ${n_1}$ is the number of variables. Thus, given the number of iterations as ${L_4}$, the overall computational complexity of Algorithm 6 can be expressed as ${\cal O}\left( {{L_4}\left( {{n_1}\ln {n_1} + {{\cal O}_1} + {{\cal O}_2}} \right)} \right)$, which is in a polynomial time. The convergence of Algorithm 6 is proved in the following Theorem 4.

\textbf{\emph{Theorem 4:}} Algorithm 6 increases the total computation rate of UEs at each iteration and finally converges within a limited number of iterations.

\emph{Proof:} Denote the objective function of problem (8) for TS with as $L_{{\rm{sum}}}^{{\rm{TS}}}$. We have
\begin{equation}
\begin{array}{l}
L_{{\rm{sum}}}^{{\rm{TS}}}\left( {{{\bf{\Lambda }} ^{(t)}},{p_i}^{(t)},f_i^{(t)},{\bf{u}}{{_{k,{\rm{TS}}}^U}^{(t)}},{\bf{u}}{{_{k,{\rm{ES}}}^D}^{(t)}}} \right)
 \le L_{{\rm{sum}}}^{{\rm{TS}}}\left( {{{\bf{\Lambda }}  ^{(t + 1)}},{p_i}^{(t)},f_i^{(t)},{\bf{u}}{{_{k,{\rm{TS}}}^U}^{(t)}},{\bf{u}}{{_{k,{\rm{ES}}}^D}^{(t)}}} \right)\\
 \;\;\;\;\;\;\;\;\;\;\;\;\;\;\;\;\;\;\;\;\;\;\;\;\;\;\;\;\;\;\;\;\;\;\;\;\;\;\;\;\;\;\;\;\;\;\;\;\;\;\;\;\;\;\;\;\;\; \le L_{{\rm{sum}}}^{{\rm{TS}}}\left( {{{\bf{\Lambda }}  ^{(t + 1)}},{p_i}^{(t + 1)},f_i^{(t + 1)},{\bf{u}}{{_{k,{\rm{TS}}}^U}^{(t)}},{\bf{u}}{{_{k,{\rm{ES}}}^D}^{(t)}}} \right)\\
 \;\;\;\;\;\;\;\;\;\;\;\;\;\;\;\;\;\;\;\;\;\;\;\;\;\;\;\;\;\;\;\;\;\;\;\;\;\;\;\;\;\;\;\;\;\;\;\;\;\;\;\;\;\;\;\;\;\; \le L_{{\rm{sum}}}^{{\rm{TS}}}\left( {{{\bf{\Lambda }}  ^{(t + 1)}},{p_i}^{(t + 1)},f_i^{(t + 1)},{\bf{u}}{{_{k,{\rm{TS}}}^U}^{(t + 1)}},{\bf{u}}{{_{k,{\rm{ES}}}^D}^{(t + 1)}}} \right).
\end{array}
\end{equation}

The first inequality comes from that ${{{\bf{\Lambda }}  ^{(t)}}}$ is solved by linear programming with fixed resource allocation $\left\{ {{p_i}^{(t)},{f_i}^{(t)}} \right\}$ and coefficient matrices $\left\{ {{\bf{u}}{{_{k,{\rm{TS}}}^U}^{(t )}},{\bf{u}}{{_{k,{\rm{ES}}}^D}^{(t )}}} \right\}$; the second and third inequalities follow the fact that $\left\{ {{p_i}^{(t + 1)},{f_i}^{(t + 1)}} \right\}$ and $\left\{ {{\bf{u}}{{_{k,{\rm{TS}}}^U}^{(t + 1)}},{\bf{u}}{{_{k,{\rm{ES}}}^D}^{(t + 1)}}} \right\}$ are optimal solutions to subproblem of UEs' resource allocation and the subproblem of coefficient matrices optimization for STAR-RIS, respectively. Considering that the total computation rate is upper-bounded, therefore, the objective function $L_{{\rm{sum}}}^{{\rm{TS}}}$ converges after several iterations.   $ \hfill{}\blacksquare $

\section{Simulation Results}
In this section, simulation results are provided to evaluate the performances of the proposed algorithms for the total computation rate maximization problems in the STAR-RIS-enhanced wireless-powered MEC system. To ensure a fair comparison, one reflecting-only RIS and one transmitting-only RIS are employed at the same location as the STAR-RIS to play as the baseline scheme (referred to as the conventional RIS) and achieve the full-space coverage \cite{XMu2021STAR}, where both the reflecting-only RIS and the transmitting-only RIS have \(M/2\) elements.
The UEs are randomly located in the reflection space and transmission space. The communication links are modeled as ${\rm{PL}} = \rho {d^{ - \alpha }}\phi $, where $\rho  =  - 30{\rm{dB}}$ is the path loss at the reference \(d_0 = 1\)m; $d$ denotes the distance between the wireless transmitter and the corresponding receiver; $\alpha$ represents the path loss factor of communication link and $\phi$ follows Rayleigh fading \cite{MSun2021}. Accordingly, we denote $\alpha _1^U$ and $\alpha _2^U$ as the path loss factors in the uplink channels from UEs to STAR-RIS and that from STAR-RIS to the AP, respectively. $\alpha _1^D$ and $\alpha _2^D$ are path loss factors in the downlink channels from STAR-RIS to UEs and that from the AP to STAR-RIS. The other simulation parameters are summarized in Table I.

%

\begin{table}[t]\small
\caption{Simulation Parameters \cite{MSun2021}}
\centering

\begin{tabular}{|c|c|c|c|}
\hline
\bfseries Parameters & \bfseries Values & \bfseries Parameters & \bfseries Values \\
\hline
Total bandwidth, $B$ &20 MHz & Noise power, ${\sigma ^2}$ & -50 dBm \\
\hline
Energy conversion efficiency, $\eta $ & 0.8 & UEs' maximum transmit power, ${P_{\max }}$ & 0.1 W \\
\hline
UEs' maximum CPU frequency, ${F_{\max }}$ & 8 GHz & Effective capacitance coefficient, $\kappa $ & ${10^{ - 28}}$ \\
\hline
The tolerant threshold, $\varepsilon ,\delta $ & ${10^{ - 4}}$  & Path loss factor, $\alpha _2^U/\alpha _1^D/\alpha _2^D$ & 3/3/3.5\\

\hline
\end{tabular}

\end{table}

\begin{figure}
\begin{minipage}[t]{0.45\linewidth}
\centering
\includegraphics[width =3.2in]{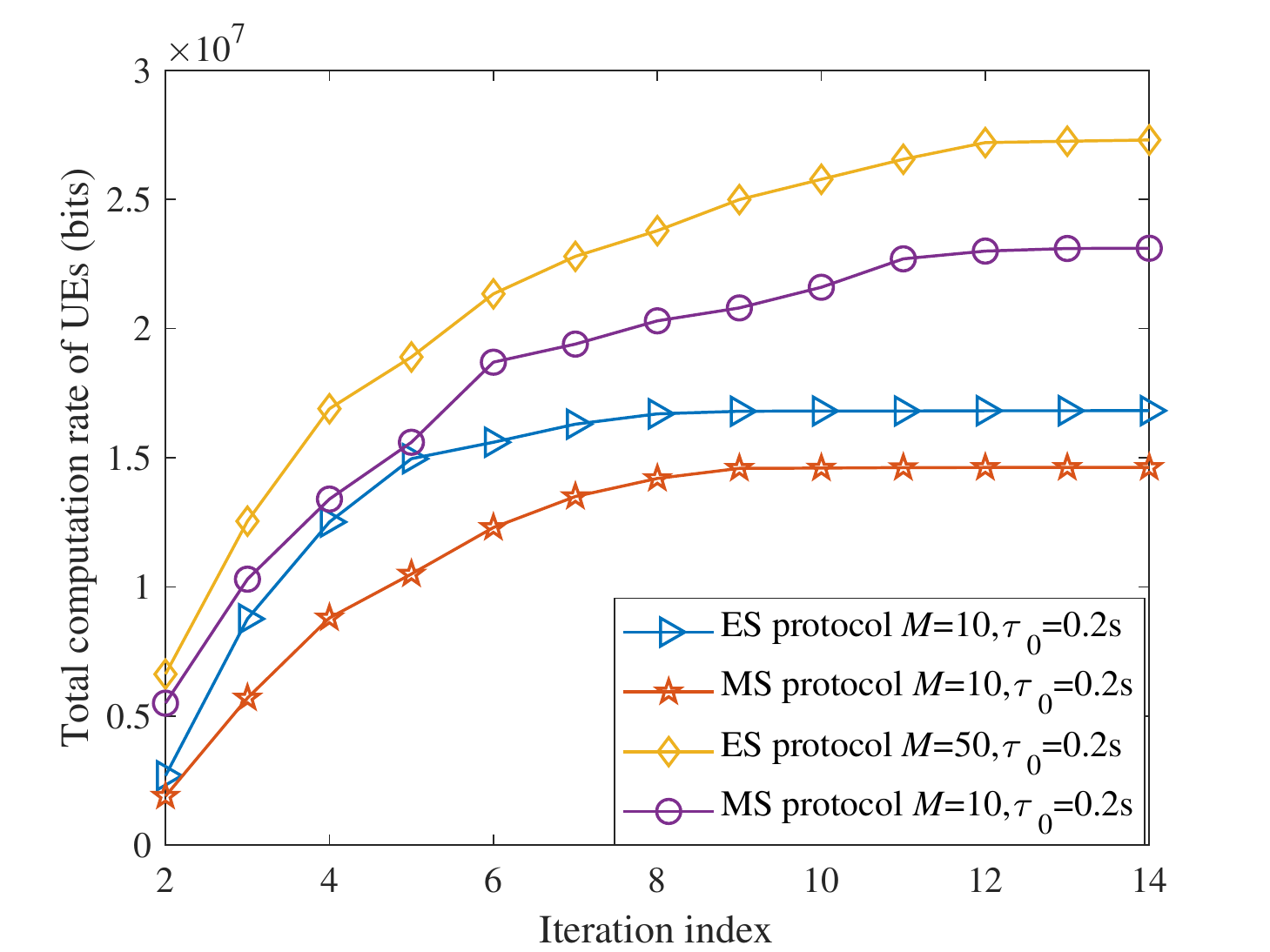}
\caption{The total computation rate of UEs versus the \protect\\ iteration index for the ES and MS protocols.} 
\end{minipage}%
\begin{minipage}[t]{0.55\linewidth}
\centering
\includegraphics[width =3.2in]{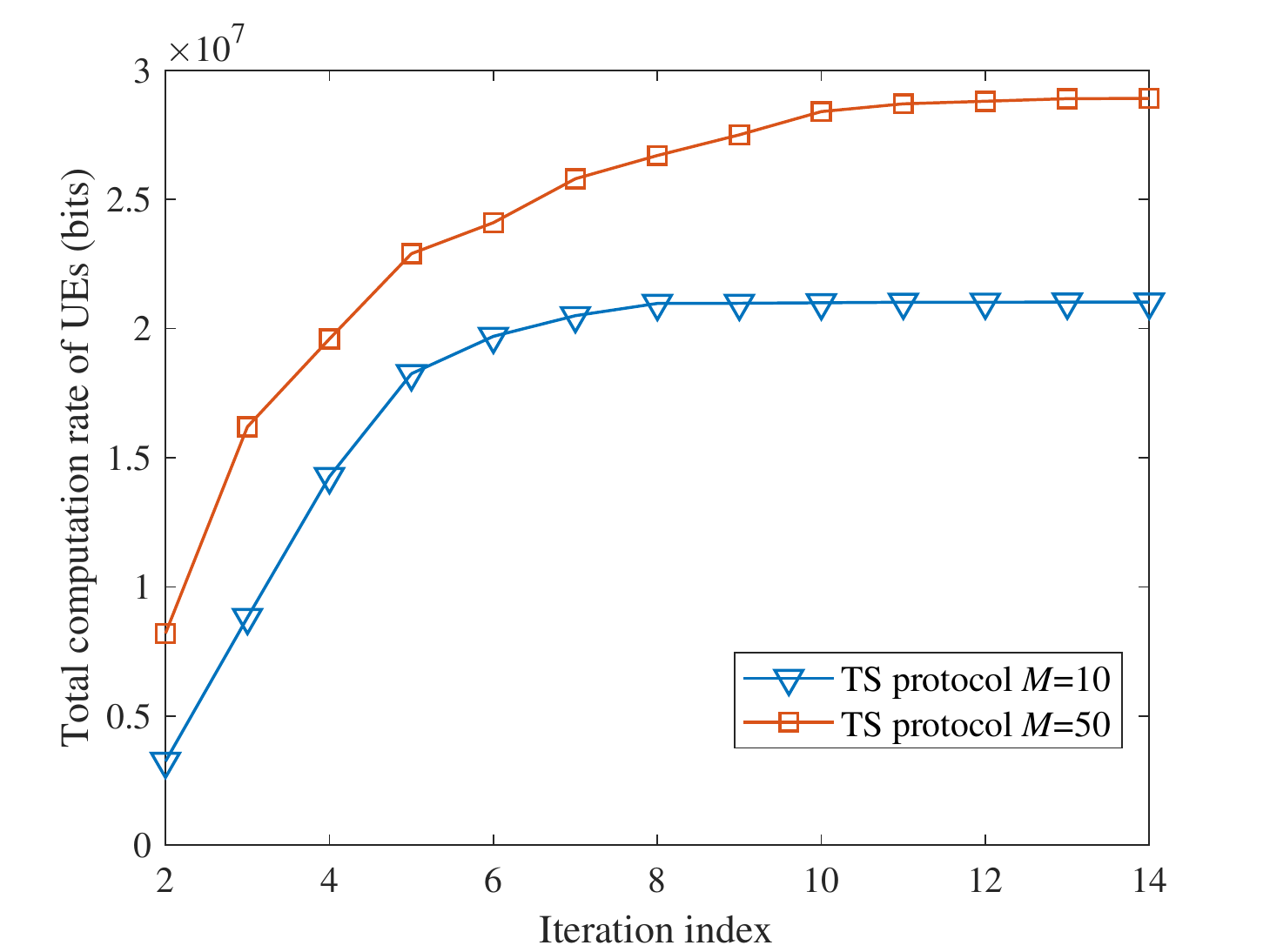}
\caption{The total computation rate of UEs versus the iteration index for the TS protocol.}
\end{minipage}

\end{figure}

%

Fig. 3 and Fig. 4 demonstrate the convergence behaviors of the proposed algorithms for ES, MS, and TS protocols, where $T = 1$s. For ES and MS protocols, the energy transfer time is given by ${\tau _0} = 0.2$s. It can be observed that the total computation rates of three protocols are monotonically increased at each iteration, and the algorithms finally converge after several iterations, which verifies the convergence analysis in Theorems 3 and 4. Besides, under different numbers of STAR-RIS elements, the proposed algorithm can still converge with a fast rate.

%
%
%

\begin{figure}
\begin{minipage}[t]{0.45\linewidth}
\centering
\includegraphics[width =3.2in]{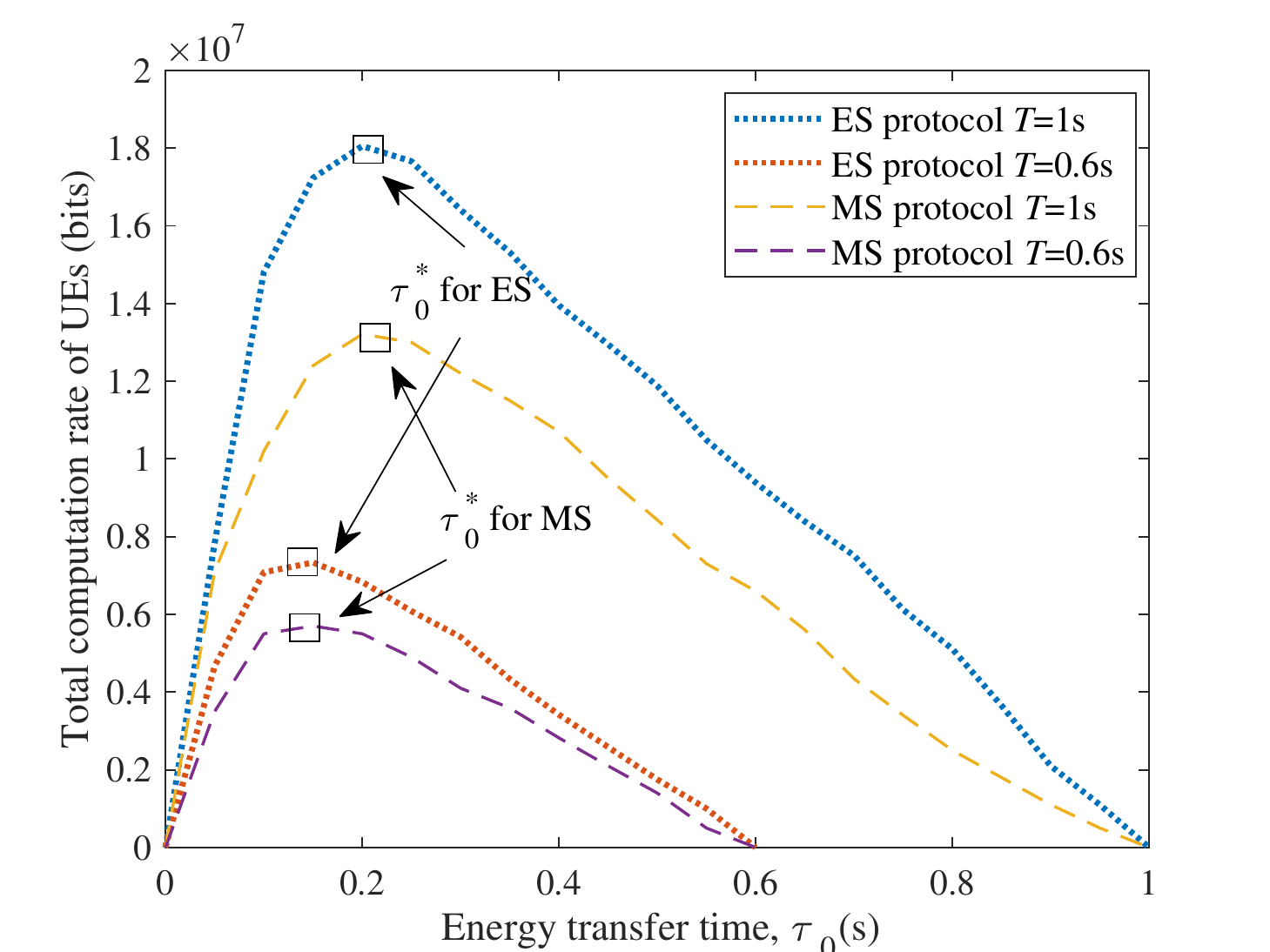}
\caption{The linear search processes of the proposed  \protect\\ Algorithm 4 for ES protocol and Algorithm 5 for  \protect\\ MS protocol.} 
\end{minipage}%
\begin{minipage}[t]{0.55\linewidth}
\centering
\includegraphics[width =3.2in]{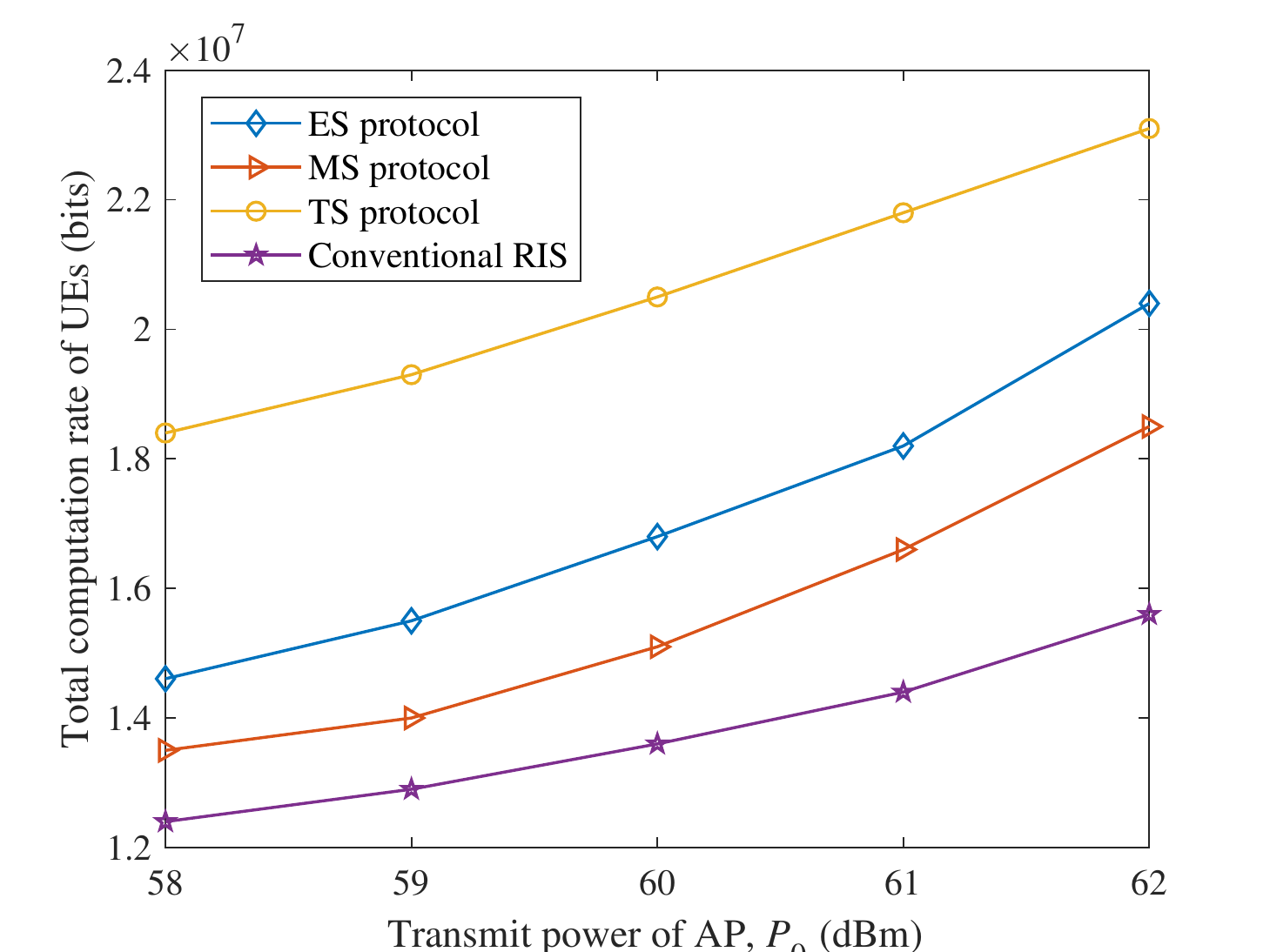}
\caption{The total computation rate of UEs versus the transmit power of the AP.}
\end{minipage}

\end{figure}

\begin{spacing}{1.56}
Fig. 5 displays the linear search processes of the proposed Algorithm 4 and Algorithm 5, where the step size  $\Delta  = 0.02$s. It is observed that for both ES and MS protocols, the total computation rate of UEs first increases as ${\tau _0}$ grows, and then starts to decrease after ${\tau _0}$ is larger than a threshold. The reasons behind this phenomenon can be explained as follows. When ${\tau _0}$ is smaller, UEs can only harvest fewer energy for task offloading and local computing. Accordingly, the total computation rate of UEs is lower. Then, with the increase of ${\tau _0}$, more energy can be harvested by UEs and thus more task bits can be offloaded to the AP or computed locally at UEs, which leads to the increase of the total computation rate. However, when ${\tau _0}$ is further increased, the rest time of the mission period used for local computing and task offloading is continuously decreased. Hence, the total computation rate of UEs is also decreased. In addition, we also observe that the optimal energy transfer time ${\tau _0^*}$ that results in the maximum total computation rate of UEs increases with the growth of the mission period. This is because with a larger mission period, UEs have enough time for local computing and task offloading. Thus, ${\tau _0^*}$ can be increased such that UEs can harvest more energy to increase the transmit power for task offloading and the CPU frequency for local computing, thereby achieving a larger computation rate.

%
%
%

In Fig. 6, the total computation rate of UEs versus the transmit power of the AP is demonstrated, where $T = 1$s and $M = 10$. As expected, the total computation rate of UEs increases as the AP's transmit power increases for all schemes. This is because when the AP transfers energy to UEs with a larger power, the UEs can harvest more energy and thus more computation tasks can be executed via task offloading and local computing. Besides, it can be observed that the conventional RIS always has the worst performance in terms of total computation rate compared with the three protocols of STAR-RIS. The reason is that compared to the conventional RIS which can only control the phase shift, the STAR-RIS has more adjustable parameters and these parameters provide extra DoFs to further enhance the offloading channel conditions of UEs.
By fully exploiting these DoFs, the STAR-RIS can always achieve higher computation rate than the conventional RIS. Moreover, regarding the three protocols for STAR-RIS, the TS protocol can achieve the best performance in comparison with the ES and MS protocols. This is because the opposite-side leakage of ES/MS protocol leads to the waste of UEs' transmit power, and hence decreases the total computation rate. As a contrast, for the TS protocol, although the offloading time is reduced due to the STAR-RIS's time allocation for reflection and transmission modes, the UEs are always served by all elements and there is no energy leaked to the opposite side of the STAR-RIS during the task offloading.
Meanwhile, if the TS protocol is employed at the STAR-RIS, the interference among UEs can be greatly reduced compared to the ES and MS protocols since the inter-user interference at the AP only comes from the UEs in the transmission/reflection space, i.e., only half-space interference exists. On the contrary, if the ES or MS protocol is employed, the inter-user interference comes from all UEs located both in the transmission space and the reflection space, i.e., there exists full-space interference. Thus, the interference is more severe at the AP and the total computation rate of UEs is decreased.
In addition, it can also be observed that the ES protocol outperforms the MS protocol since the MS protocol is a special case of ES protocol from the mathematical perspective, as we state in problem (7).

%
%
%

\begin{figure}
\begin{minipage}[t]{0.45\linewidth}
\centering
\includegraphics[width =3.2in]{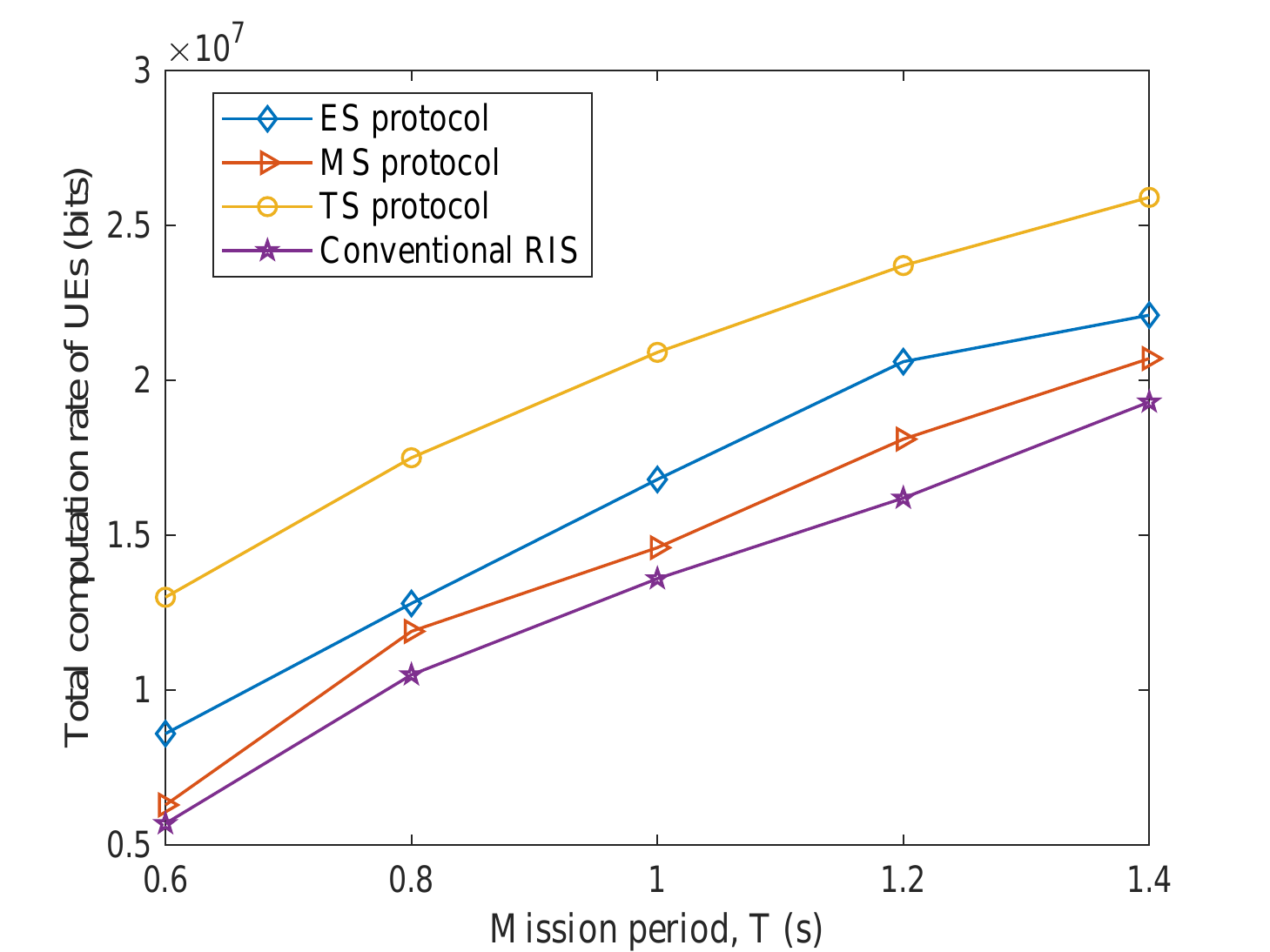}
\caption{The total computation rate of UEs versus \protect\\ the mission period.} 
\end{minipage}%
\begin{minipage}[t]{0.55\linewidth}
\centering
\includegraphics[width =3.2in]{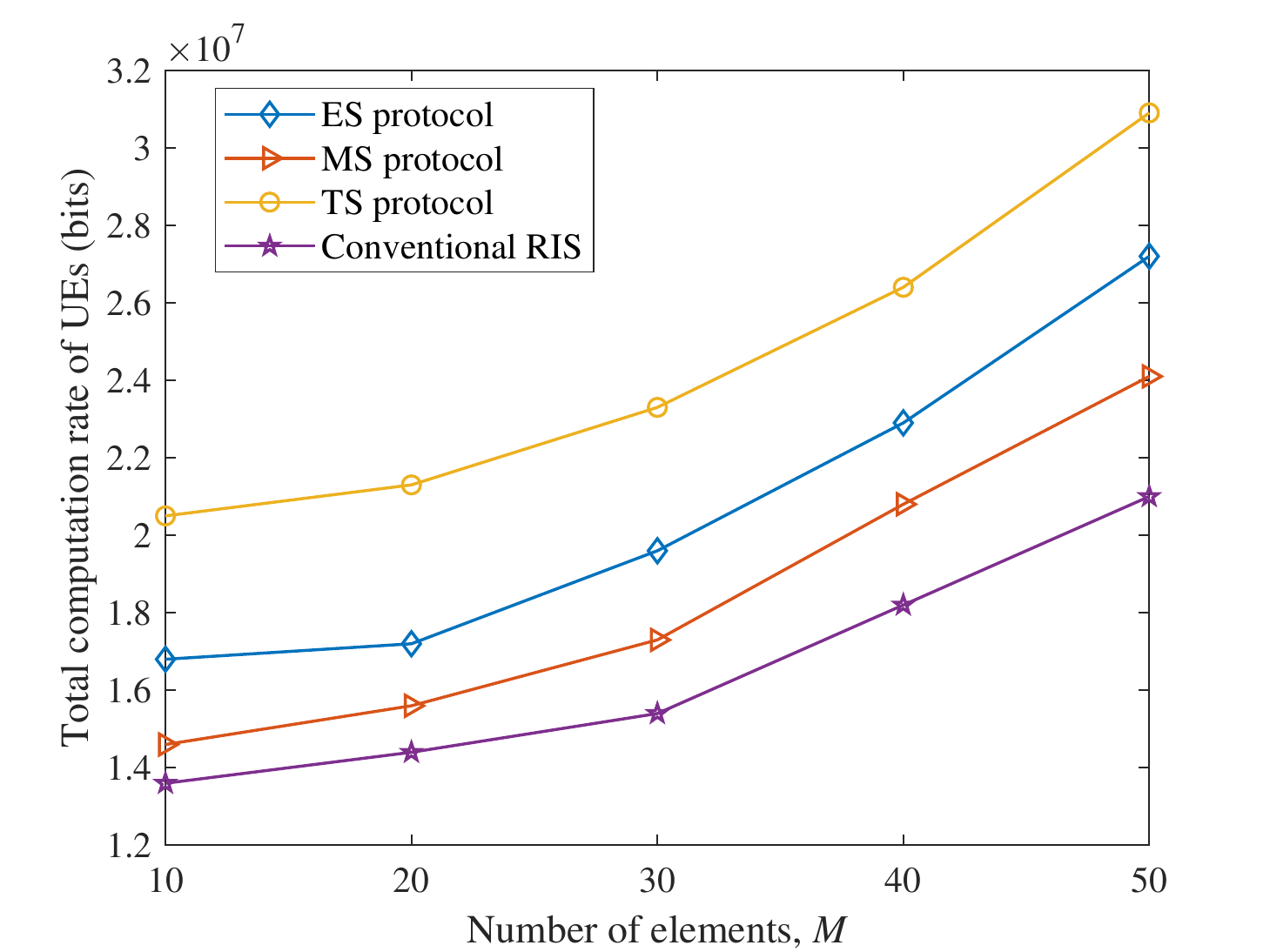}
\caption{The total computation rate of UEs versus the number of RIS elements.}
\end{minipage}

\end{figure}

Fig. 7 shows the impact of the mission period $T$ on the total computation rate of UEs.
It is observed that the total computation rate increases with the growth of the mission period. This is because when the mission period increases, the energy transfer time and the energy harvested by UEs also increase, which provides more energy supplies for UEs to execute more task bits. Besides, with a larger mission period, the task offloading and local computing can be executed with longer time, which further increases the total computation rate of UEs. It is also observed that the STAR-RIS outperforms the conventional RIS and the TS has the best performance, which coincides with the results shown in Fig. 6.
\end{spacing}
%
%
%

Fig. 8 presents the total computation rate of UEs versus the number of RIS's elements.
It can be found that the total computation rate of UEs increases with the number of RIS's elements. Moreover, we also observe that the performance gap between the STAR-RIS and conventional RIS becomes larger as the number of RIS's elements increases. The reason is that when the number of RIS elements increases, the additional elements can provide more opportunities for designing more efficient configuration strategy of STAR-RIS, and thus a higher performance gain can be achieved.

%
%

\begin{figure}
\begin{minipage}[t]{0.45\linewidth}
\centering
\includegraphics[width =3.2in]{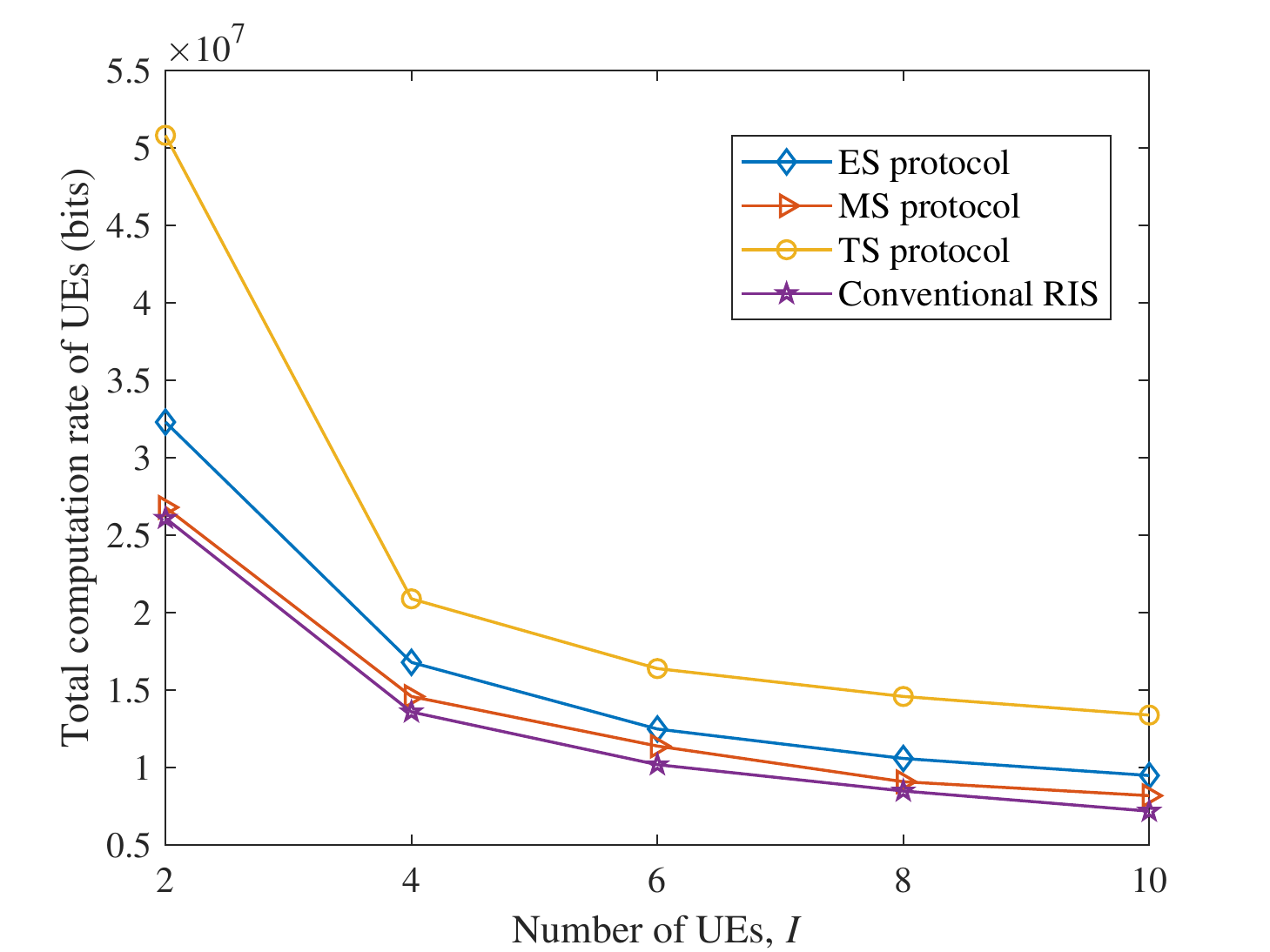}
\caption{The total computation rate of UEs versus \protect\\ the number of UEs.} 
\end{minipage}%
\begin{minipage}[t]{0.55\linewidth}
\centering
\includegraphics[width =3.2in]{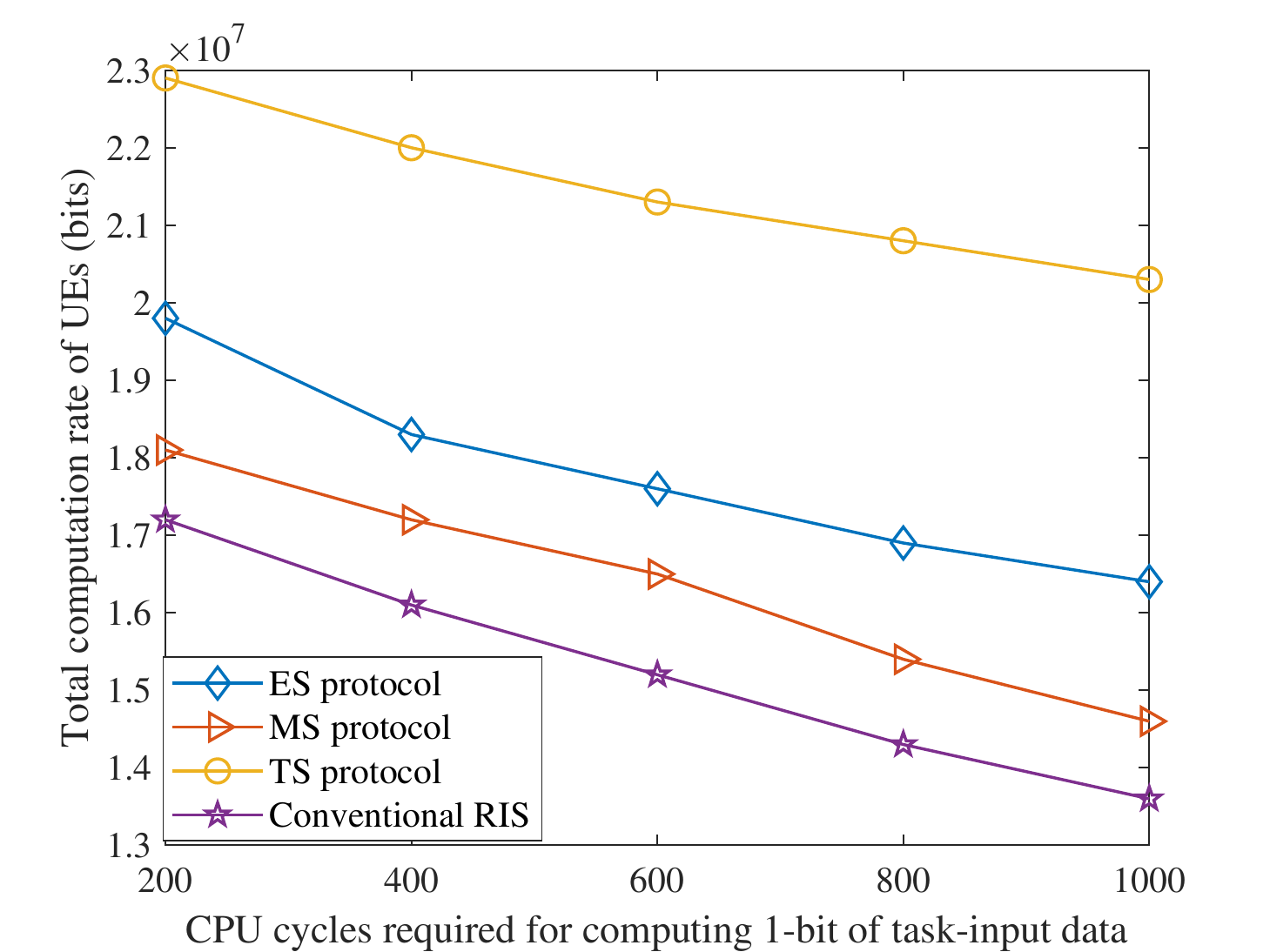}
\caption{The total computation rate of UEs versus the CPU cycles required for computing 1-bit of task-input data.}
\end{minipage}

\end{figure}

In Fig. 9, the total computation rate of UEs is shown against the number of UEs.
To ensure a fair comparison, there are the same number of UEs in the transmission space and reflection space. From Fig. 8, it can be found that the total computation rate of UEs decreases with the increase of the number of UEs. The reason is that when the UEs perform task offloading, the NOMA protocol is applied, which allows all UEs to access the AP and offload task bits at the same time and frequency. Thus, the inter-user interference becomes more severe with larger number of UEs, which leads to the degradation of the total computation rate.


%
In Fig. 10, we study the impacts of the CPU cycles required for computing 1-bit of task-input data (i.e., ${C_i}$) on the total computation rate 
of UEs.
It is observed that the total computation rate decreases with the increase of ${C_i}$.
Apparently, when ${C_i}$ increases, the energy consumed for executing 1-bit task data via local computing will be increased. Thus, with the limited energy harvested from the AP, the amount of task bits computed locally by UEs decreases. Accordingly, the total computation rate of UEs is also decreased.

\section{Conclusions}
In this paper, the STAR-RIS-enhanced wireless-powered MEC system has been investigated, where the STAR-RIS was deployed to assist UEs' task offloading and AP's energy transfer. The total computation rate of UEs was maximized by jointly optimizing the energy transfer time, transmit power and CPU frequencies of UEs, and the configuration design of STAR-RIS. Three operating protocols of STAR-RIS were considered during the task offloading. To solve the formulated non-convex problems, based on the penalty method, the SCA technique and the linear search method, an iterative algorithm was proposed to solve the ES problem. Then, the proposed algorithm for ES protocol was extended to solve the MS and TS problems. Simulation results revealed that the STAR-RIS outperformed the traditional reflecting/transmitting-only RIS. More importantly, the TS protocol can achieve the largest computation rate among the three operating protocols of STAR-RIS.

%
%

\bibliographystyle{IEEEtran}
\bibliography{IEEEabrv,bib2014}

\end{sloppypar}
\end{document}